\documentclass[fleqn,usenatbib,referee]{mnras}
\usepackage[dvipdfmx]{graphicx}
\usepackage{pdfpages}
\usepackage{amsmath}  
\usepackage{amssymb}  
\usepackage[T1]{fontenc}
\usepackage{bm}
\usepackage{color}
\DeclareRobustCommand{\VAN}[3]{#2}
\let\VANthebibliography\thebibliography
\def\thebibliography{\DeclareRobustCommand{\VAN}[3]{##3}\VANthebibliography}
\newcommand{\msun}{{\rm M}_\odot}
\newcommand{\cc}{{\rm cm^{-3}}}
\newcommand{\gcm}{{\rm g\,cm^{-3}}}

\title[Dust motion in a Disk ]{Dust Motion and Possibility of Dust Growth in a Growing Circumstellar Disk}

\author[Koga \& Machida]{
Shunta Koga$^{1}$\thanks{E-mail: kshunta3403@gmail.com} and Masahiro N. Machida$^{1}$\thanks{E-mail: machida.masahiro.018@m.kyushu-u.ac.jp}
\\
$^{1}$Department of Earth and Planetary Sciences, Faculty of Sciences, Kyushu University, Fukuoka 819-0395, Japan\\
}
\begin{document}
\label{firstpage}
\pagerange{\pageref{firstpage}--\pageref{lastpage}}
\maketitle
\begin{abstract}
We calculate the evolution of a star-forming cloud core using a three-dimensional resistive magnetohydrodynamics simulation, treating dust grains as Lagrangian particles, to investigate the dust motion in the early star formation stage. 
We prepare six different-sized set of dust particles in the range $a_{\rm d}=0.01$--$1000\,\mu$m, where $a_{\rm d}$ is the dust grain size. 
In a gravitationally collapsing cloud,  a circumstellar disk forms around a protostar and drives a protostellar outflow. 
Almost all the small dust grains ($a_{\rm d} \lesssim 10$--$100\,\mu$m) initially distributed in the region $\theta_0 \lesssim 45^\circ$ are ejected from the center by the outflow, where $\theta_0$ is the initial zenith angle relative to the rotation axis,  whereas only a small number of the large dust grains ($a_{\rm d} \gtrsim 100\,\mu$m) distributed in the region are ejected. 
All other grains fall onto either the protostar or disk without being ejected by the outflow.
Regardless of the dust grain size, the behavior of the dust motion is divided into two trends after dust particles settle into the circumstellar disk.  
The dust grains reaching the inner disk region from the upper envelope preferentially fall onto the protostar, 
while those reaching the outer disk region or disk outer edge from the envelope can survive without an inward radial drift. 
These surviving grains can induce dust growth. 
Thus, we expect that the outer disk regions could be a favored place of planet formation. 
\end{abstract}
\begin{keywords}
stars: formation --
stars: magnetic field -- 
MHD -- 
ISM: dust, , extinction-- 
ISM: jets and outflows
\end{keywords}
\section{Introduction}
Many planetary systems have been confirmed since 1995 \citep{Mayor1995}. 
Planets form in disks around young stars \citep{2021ApJ...916L...2B}, 
as shown in recent ALMA observations of the formation sites of planets \citep[e.g.][]{2015ApJ...808L...3A}. 
Ring and gap structures can be  clearly seen in protoplanetary disks observed by dust thermal emission \citep{Andrews2018}, 
and these structures are likely to be related to planet formation \citep{2020ARA&A..58..483A}. 
It is considered that dust grains aggregate and grow to form planets in the disks around young stars \citep{1985prpl.conf.1100H}. 
However, it is very difficult to observe the disks in the very early phase during which  planet formation begins.
Therefore, we cannot confidently identify when and how dust  grows and planet formation begins in a disk based only on observations. 

Theoretical studies can help us to understand dust growth and  planet formation \citep[e.g.][]{2008ARA&A..46...21B}. 
In such studies,  planet formation (or dust growth) is conventionally considered to  begin in an isolated and relatively low-mass disk after the gas accretion onto the disk ends. 
Such a disk is called  the minimum mass solar nebula (MMSN, \citealt{1985prpl.conf.1100H}) and is observed around Class II young stellar objects.   
However, recent observations have confirmed signs of planet formation in very young growing circumstellar disks around Class 0 and I protostars \citep{2020ApJ...902..141S,2020A&A...640A..19T,2020A&A...642L...7P}, implying that planet formation begins earlier than previously thought.

The formation and evolution of disks around young stars have been theoretically investigated in the framework of star formation \citep[e.g.][]{2011MNRAS.413.2767M}. 
Physical quantities such as the mass and radius of the disk vary over time in the star formation process \citep{2020ApJ...896..158T, 2021MNRAS.502.4911X, 2021A&A...648A.101L}. 
Thus, it is not sufficient to consider the MMSN to be the initial condition  of planet formation. 
It is appropriate to consider the dust growth and planet formation from the beginning of the disk formation epoch in the star formation process.

The star formation process has been investigated with three-dimensional magneto hydrodynamics (MHD) simulations, in which only the gas dynamics are calculated and the dust motion is not explicitly considered \citep{2004MNRAS.348L...1M, 2006ApJ...641..949B, 2008A&A...477....9H, 2013ApJ...763....6T, 2014MNRAS.438.2278M,2015ApJ...810L..26T, 2021MNRAS.508.2142X}.   
However, it is important to consider the behavior of dust grains in the star formation process to investigate the dust growth and planet formation, because  dust is a fundamental element of planet formation.
There are a few studies calculating the dust motion  in three-dimensional MHD star-formation simulations \citep{2020A&A...641A.112L, 2021ApJ...920L..35T, 2022arXiv220712907K}.
\citet{2020A&A...641A.112L} performed a star formation simulation with the inclusion of dust and detailed the increase and decrease of dust around the circumstellar disk \citep[see also][]{2021ApJ...913..148T}.
\citet{2021ApJ...920L..35T} also calculated the star formation process including dust and showed that the dust grains are swept by the protostellar outflow and  that part of the swept dust falls onto the disk. 
They considered the dust growth or dust size evolution with single-sized approximation. 
It should be noted that, recently, the dust growth in the star formation process were also investigated with various (or different) approaches in \citet{2022A&A...666A..27M,2021A&A...649A..50M}, \citet{2022MNRAS.514.2145B} and \citet{2022MNRAS.515.4780T}.  

Our calculation settings are approximately similar to \citet{2020A&A...641A.112L} and \citet{2021ApJ...920L..35T}. 
Our previous study \citep[][hereafter Paper I]{2022arXiv220712907K} showed almost the same results as in \citet{2020A&A...641A.112L} and \citet{2021ApJ...920L..35T}, but used a novel method in which the behavior of the dust grains is calculated using a Lagrangian formulation and the fluid motion is calculated using an Eulerian formulation. 
Following Paper I,  this study focuses on the behavior of dust in both a gravitationally collapsing cloud and a circumstellar disk, starting the simulation from the prestellar stage (or the molecular cloud core). 
Our study differs from the past studies of \citet{2020A&A...641A.112L} and \citet{2021ApJ...920L..35T} in the treatment of dust, 
in that while past studies adopted a fluid approximation, our study treats the dust particles as Lagrange particles, calculated separately from the gas component (for detail, see Paper I).
Using this treatment, we can individually trace the motion of each dust particle and investigate the behavior of dust in the disk during the early star formation stage. 
As described above, the implementation of an MHD nested grid code in the treatment of the dust was presented in Paper I, in which we also presented the trajectories of dust in the infalling envelope focusing on the coupling between the gas and dust.
In this paper,  we mainly present the motion of dust in the circumstellar disk.

The structure of this paper is as follows. 
The numerical settings and methods are described in \S2 and the calculation results are presented in \S3.
We discuss dust motion, dust growth and planet formation in the early star formation stage in \S4. 
A summary is presented in \S5.

\section{Numerical method and settings} 
The numerical method and settings are the same as in Paper I 
and we only briefly explain them here (for details,  see \S2 and Appendix of Paper I).

As the initial state, a gas sphere with a critical Bonner--Ebert density profile is adopted.  
The mass and radius of the initial cloud are $M_{\rm cl}=1.25\,\msun$ and $R_{\rm cl}=6.13\times10^3$\,au, respectively. 
A rigid rotation ($\Omega_0=2\times10^{-13}$\,s$^{-1}$) is added to the initial cloud and a uniform magnetic field ($B_0=5.1\times10^{-5}$\,G) is imposed over the whole computation domain. 
The mass-to-flux ratio normalized by the critical value $(2\pi G^{1/2})^{-1}$ is $\mu=3$. 
The initial state is the same as that in \citet{2017ApJ...835L..11T} and \citet{2020ApJ...905..174A}. 

In the simulation, the MHD part is calculated by our three-dimensional non-ideal MHD nested grid code \citep{2004MNRAS.348L...1M, 2006ApJ...645.1227M, 2009ApJ...699L.157M,2013MNRAS.431.1719M}. 
The spatial resolution and generation criterion of the grid are described in Paper I. 
The cell width and grid size of the finest grid are $0.374$\,au and $24$\,au, respectively, while those of the coarsest grid are $3.07\times10^3$\,au and $1.96\times10^5$\,au. 
The number of cells in each grid is ($i$, $j$, $k$) = (64, 64, 64).
Five levels of grid are prepared  before the calculation, while 14 levels of grid are nested just before protostar formation (for details, see Paper I). 
In the simulation, a sink of radius $r_{\rm sink}=1$\,au is created after the density exceeds $n_{\rm thr}=10^{13}\,\cc$ \citep{2014MNRAS.438.2278M,2016MNRAS.463.4246M}. 
In addition, dust particles are implemented as super-particles, for which  the feedback from the dust to the gas is ignored, as described in Paper I.
It should be noted that we assume that the dust particles do not interact with each other.

We include a total of 102,984 dust particles in the simulation.
The initial  spatial distribution of the particles is shown in Table~\ref{table:dustdistribution}, in spherical coordinates. 
As described in the table, the dust locations are distributed every 10\,au in the range 10--6130\,au in the radial direction $r_0$,
every 90$^\circ$ in the azimuthal direction $\phi_0$,  and every 15$^\circ$ in the zenith direction $\theta_0$. 
We prepare six different grain sizes (or six different-size set of grains) in the range $a_{\rm d}=$0.01--1000\,${\rm \mu m}$,  listed in Table~\ref{table:dustsize}, where $a_{\rm d}$ is the radius of a dust grain. 
The settings for the distribution of dust particles adopted in this study are identical to those in Paper I. 

\begin{table}
 \caption{Initial spatial distributions of dust particles}
 \label{table:dustdistribution}
 \centering
  \begin{tabular}{cl}
   \hline
   Coordinate & Initial particle locations \\
   \hline \hline
   $r_0$ & 10--6130\,au \ (every 10\,au, 613 locations) \\
   $\phi_0$ & 0$^\circ$, 90$^\circ$, 180$^\circ$, 270$^\circ$ \ (every 90$^\circ$, 4 locations) \\
   $\theta_0$ & 0$^\circ$, 15$^\circ$, 30$^\circ$, 45$^\circ$, 60$^\circ$, 75$^\circ$, 90$^\circ$ \ (every 15$^\circ$, 7 locations) \\
   \hline
  \end{tabular}
\end{table}

\begin{table}
 \caption{Dust grain sizes used in the calculation}
 \label{table:dustsize}
 \centering
  \begin{tabular}{cl}
   \hline
   Dust grain size $a_{\rm d}$ [$\rm \mu$m] \\
   \hline \hline
   0.01, 0.1, 1, 10, 100, 1000  \\
   \hline
  \end{tabular}
\end{table}

\section{Results}
\label{sec:results}
\subsection{Time sequence of gas density and velocity distributions}
\label{subsec:gasevol}
Before presenting the dust motion results, we will describe the gas evolution.
Since  the dust grains are coupled with the gas to some degree, it is important to examine the gas evolution from a Eulerian perspective. 

\begin{figure*}
    \centering
    \includegraphics[width=0.8\linewidth]{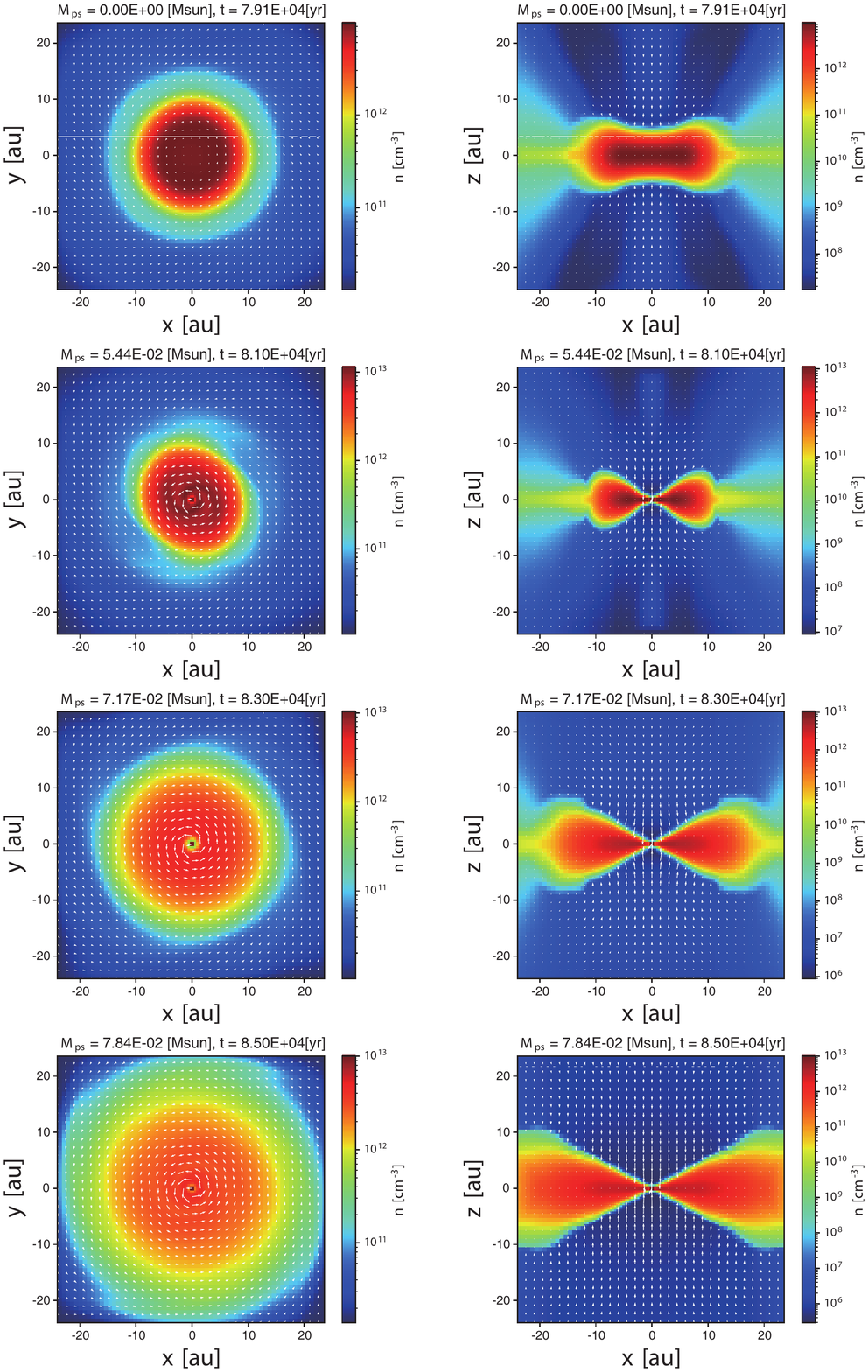}
    \caption{Gas density (color) and velocity (arrows) distributions on the $z=0$ (left) and $y=0$ (right) planes at four different epochs. The protostellar mass (or sink mass) and elapsed time are listed above each panel.}
    \label{fig:gevol}
\end{figure*}

Fig.~\ref{fig:gevol} plots the density and velocity distributions of the gas at four different epochs. 
An elliptical structure, which corresponds to the first core supported by both thermal pressure and rotation, can be seen just before protostar formation (top right panel of Fig.~\ref{fig:gevol}). 
After protostar formation,  the central region shrinks (right panels of Fig.~\ref{fig:gevol}) and a rotationally supported disk appears (left panels of Fig.~\ref{fig:gevol}). 
Since gas accretion onto the disk continues during the simulation, the disk increases in size with time. 
The simulation was stopped at $t =$ 85000\,yr after the cloud collapse begins.
At the end of the simulation, the protostellar mass (or sink mass) is $M_{\rm ps} = 0.0784\,\msun$, which is 6.3 $\%$ of the initial cloud mass.
At this epoch, the disk has a radius of about $\sim20$\,au (bottom left panel of Fig.~\ref{fig:gevol}). 

\begin{figure*}
    \centering
    \includegraphics[width=0.8\linewidth]{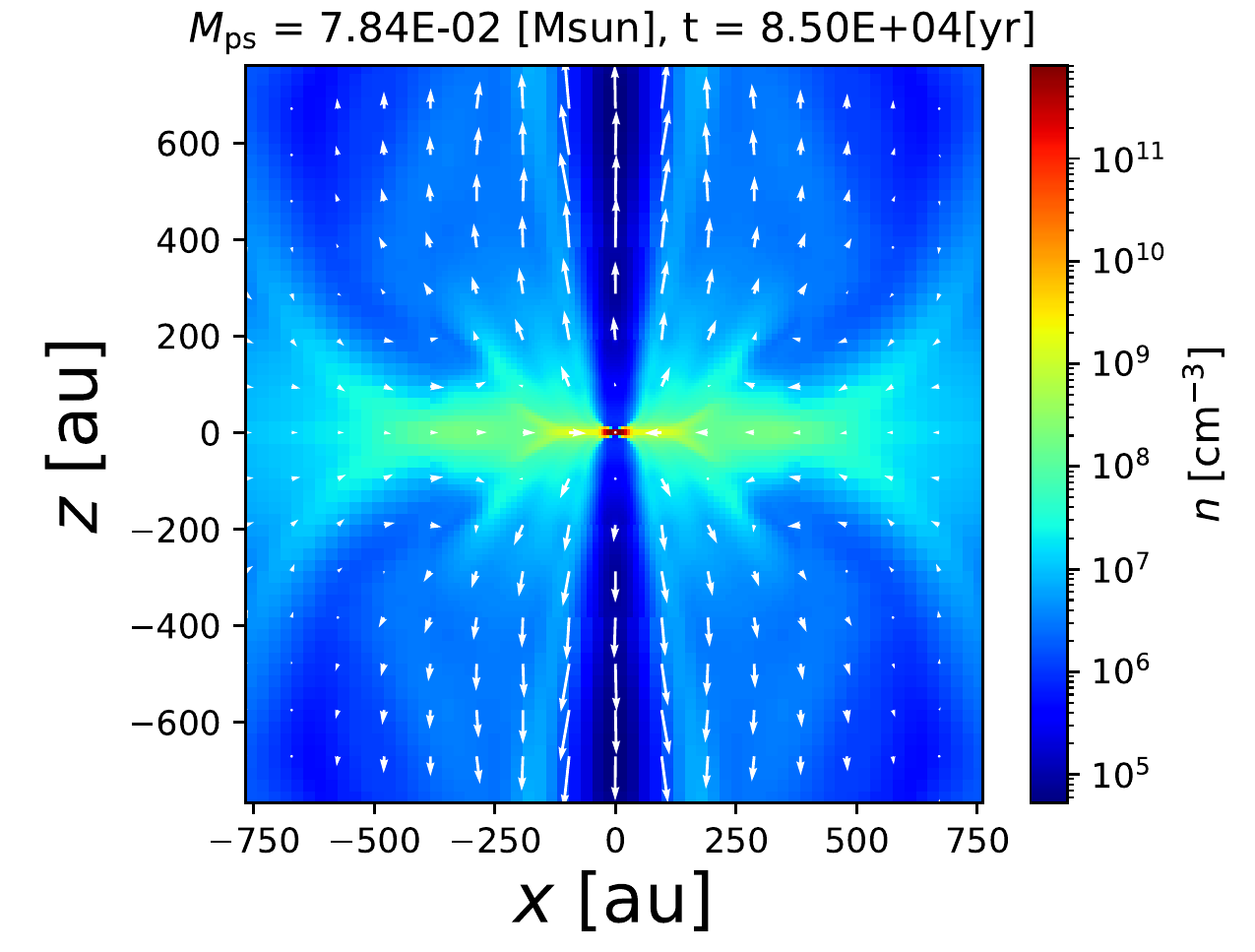}
    \caption{Gas density (color) and velocity (arrows) distributions on the $y=0$ plane. The protostellar mass and the elapsed time after the cloud collapse begins are given above the plot.}
    \label{fig:outflow}
\end{figure*}

Fig.~\ref{fig:outflow} shows the density and velocity distributions at the same epoch as in the bottom panels of Fig.~\ref{fig:gevol},
though note that the spatial scales of Fig.~\ref{fig:gevol} ($\sim50$\,au) and Fig.~\ref{fig:outflow} ($\sim1500$\,au) are very different.  
Fig.~\ref{fig:outflow} indicates that the protostellar outflow is driven near the center of the cloud. 
At the end of the calculation, the size of the outflow exceeds $\sim1000$\,au. 
As described below, a portion of the dust grains is ejected from the central region or disk by the outflow (see Paper I). 

\subsection{Motion of dust particles}
\label{subsec:dustr}
\subsubsection{Dust motion from envelope to disk}
\label{subsub:envelope}
As described in \S \ref{subsec:gasevol}, the gas evolution  is obtained from a Eulerian simulation  
while the dust grains are calculated as Lagrange particles. 
Thus, we can trace the trajectory of each dust particle with the gas evolution. 
In this subsection, we show the dependence of the dust orbital evolution on their initial positions. 

\begin{figure*}
\begin{tabular}{cc}
\begin{minipage}{0.4\hsize}
\includegraphics[width=\columnwidth]{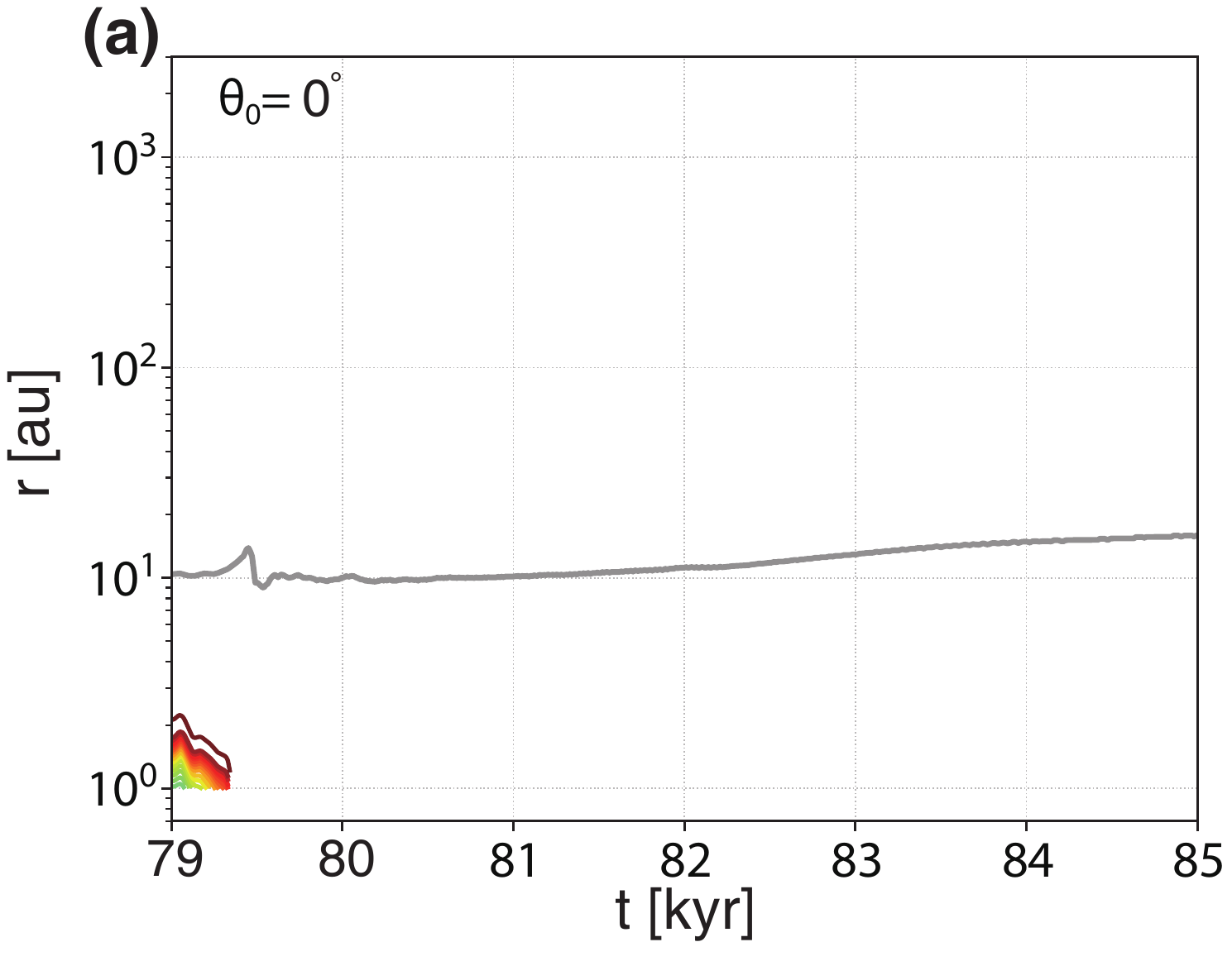}
\end{minipage}
\begin{minipage}{0.4\hsize}
\includegraphics[width=\columnwidth]{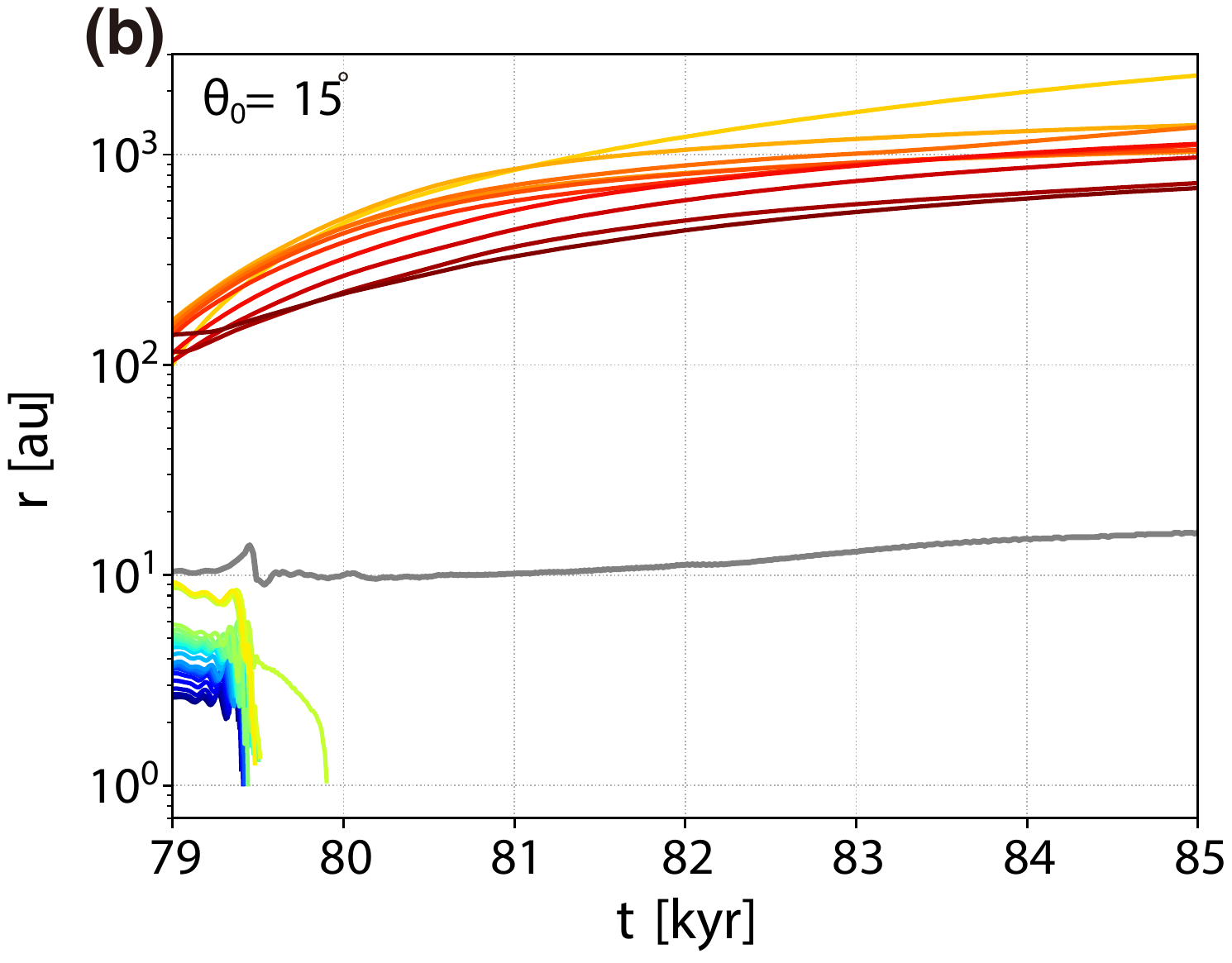}
\end{minipage}
\\
\\
\begin{minipage}{0.4\hsize}
\includegraphics[width=\columnwidth]{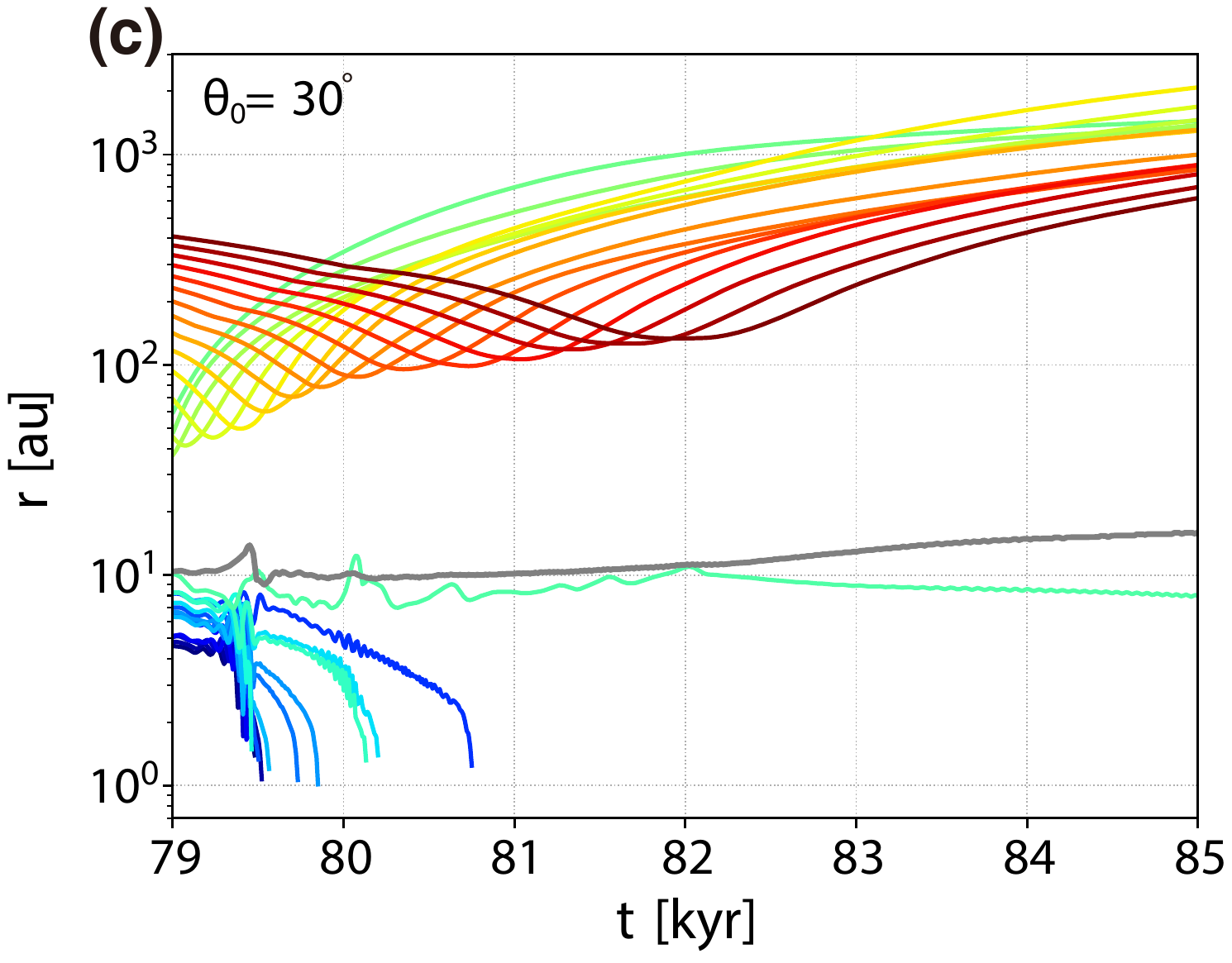}
\end{minipage}
\begin{minipage}{0.4\hsize}
\includegraphics[width=\columnwidth]{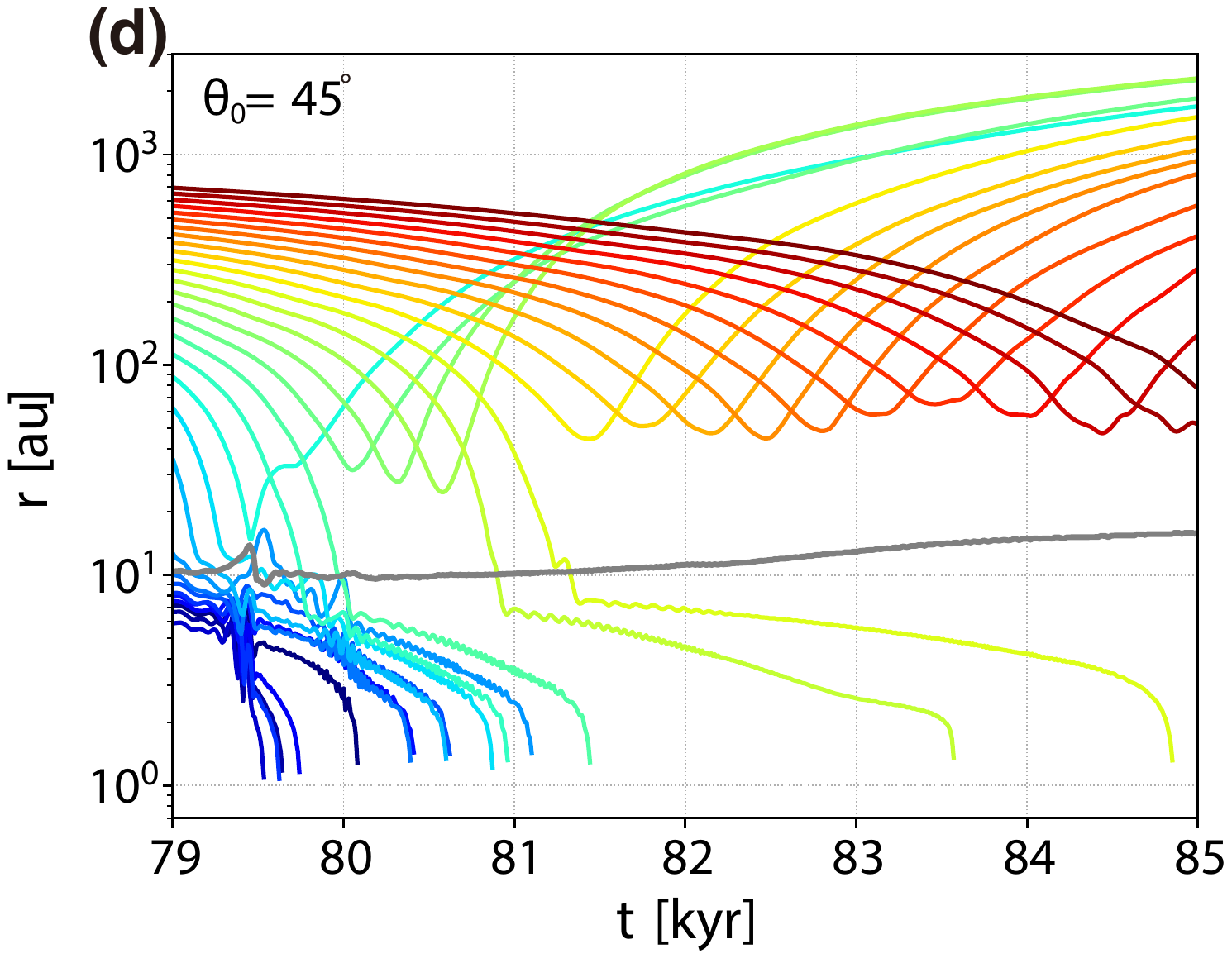}
\end{minipage}
\\
\\
\begin{minipage}{0.4\hsize}
\includegraphics[width=\columnwidth]{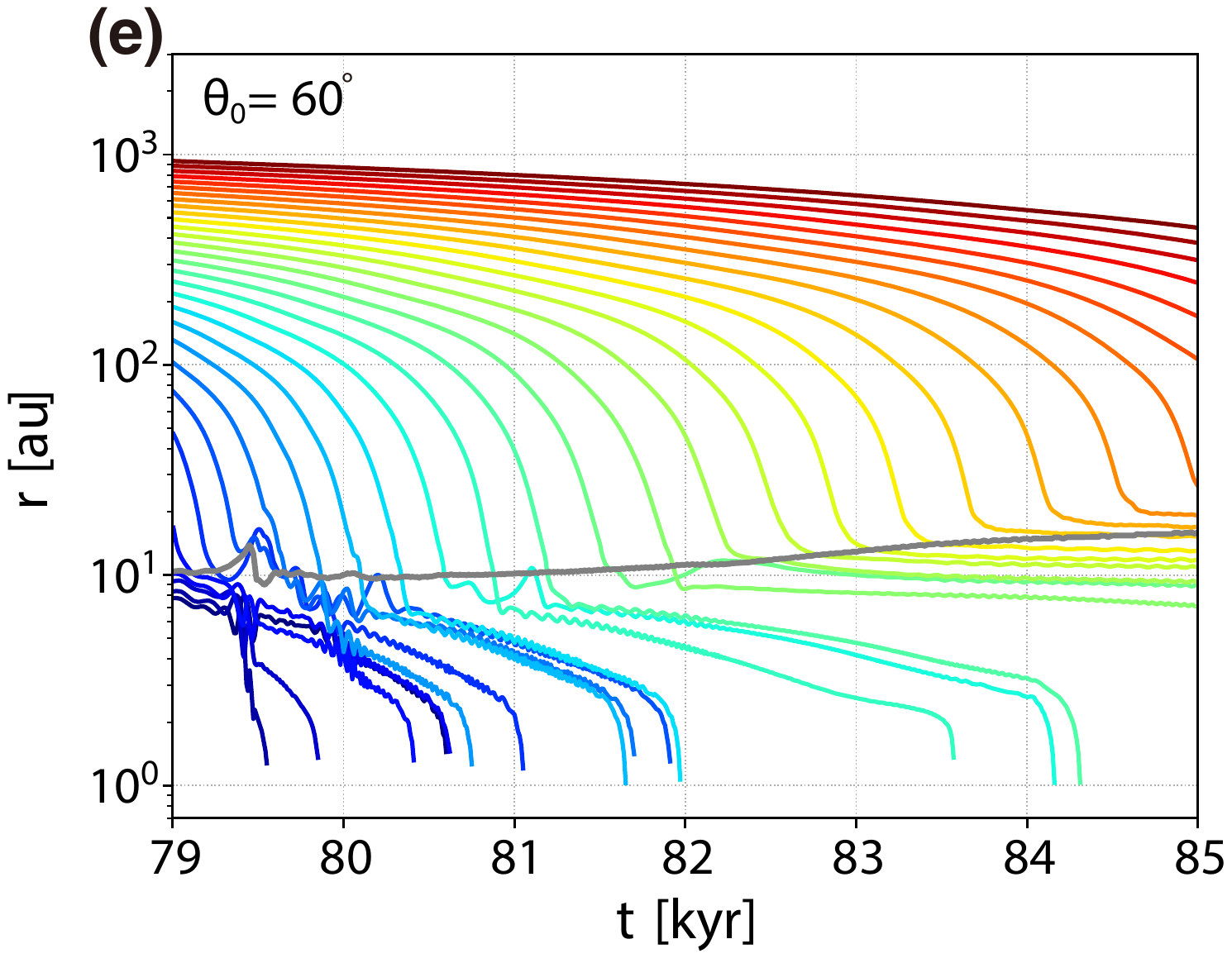}
\end{minipage}
\begin{minipage}{0.4\hsize}
\includegraphics[width=\columnwidth]{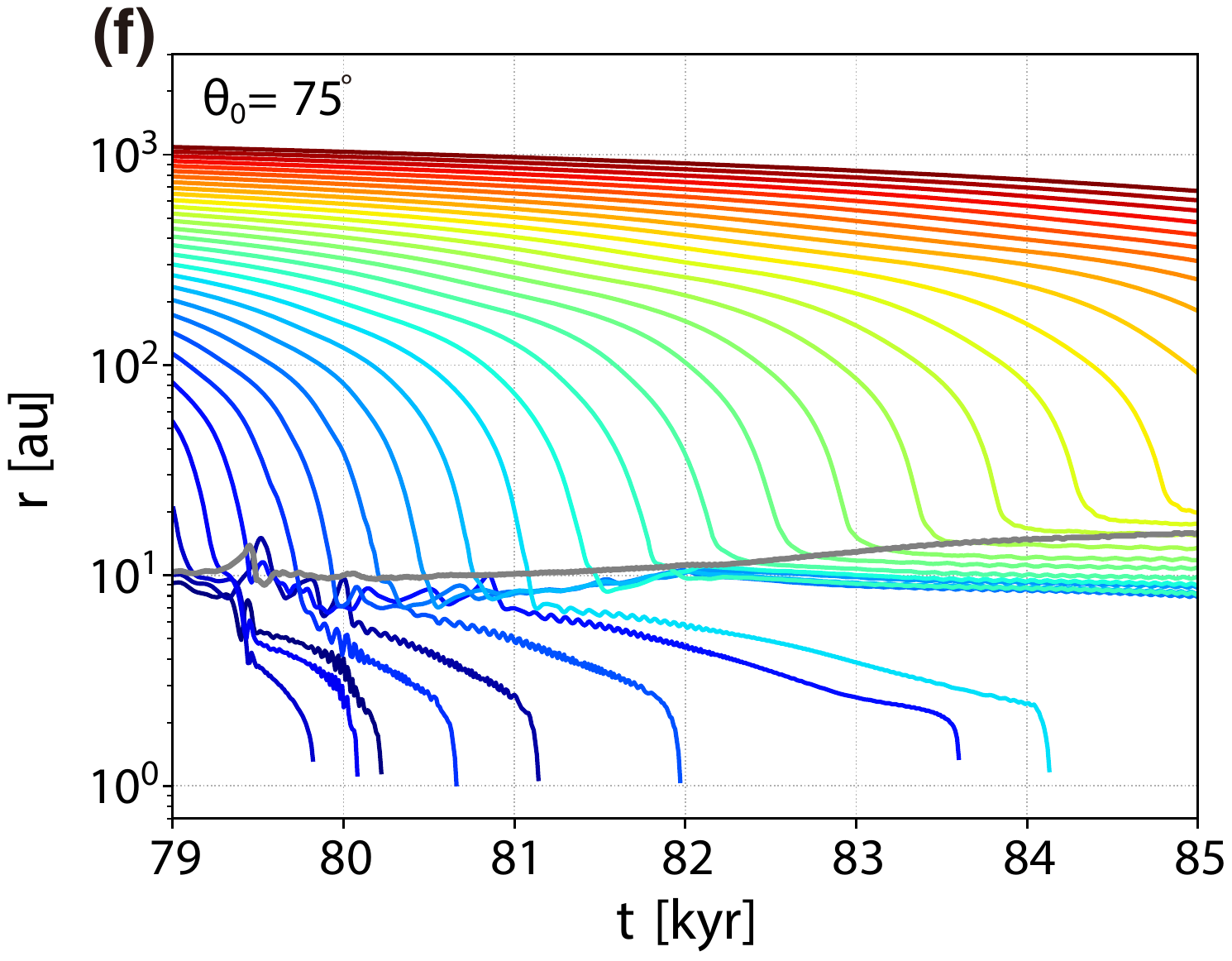}
\end{minipage}
\\
\\
\begin{minipage}{0.4\hsize}
\includegraphics[width=\columnwidth]{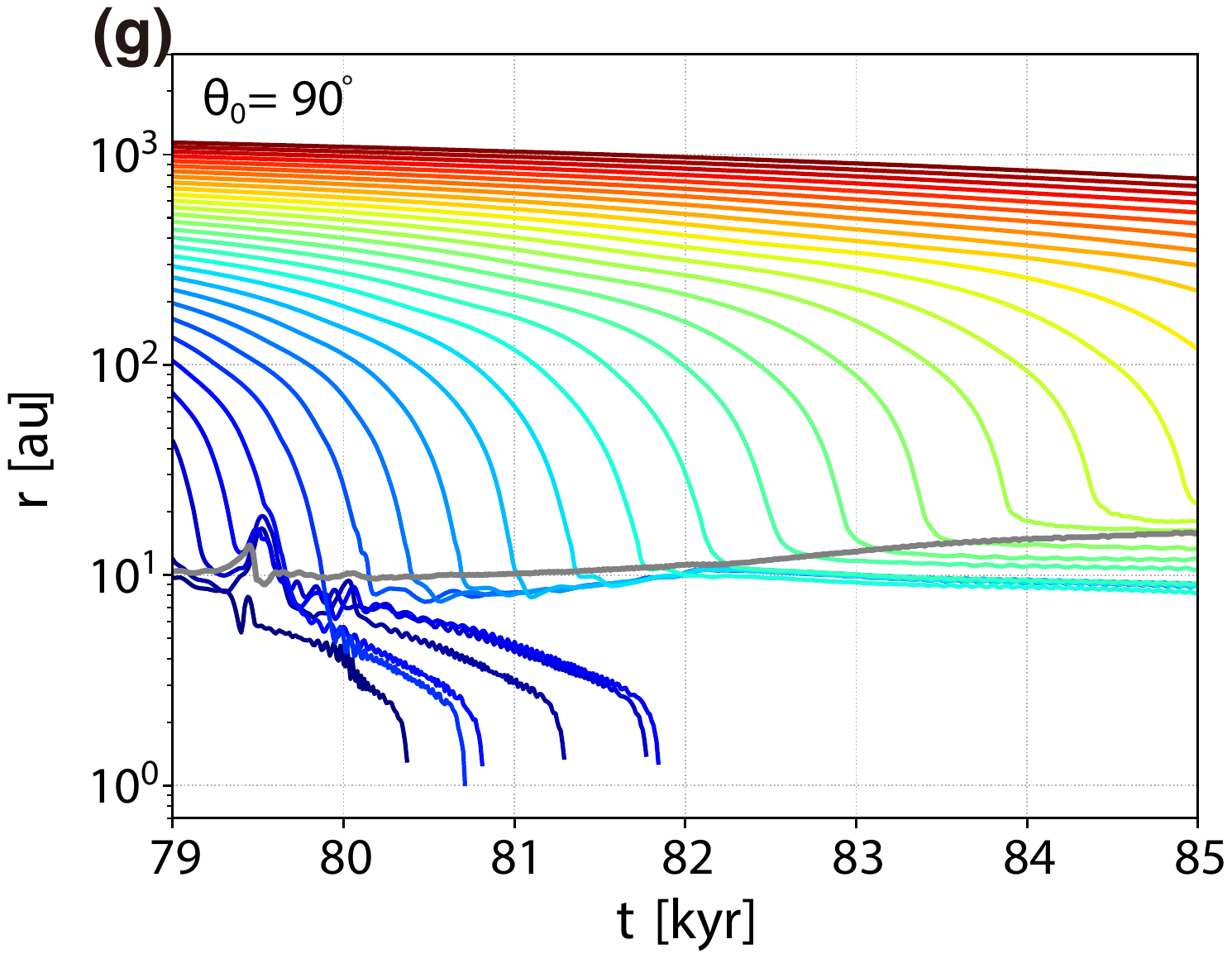}
\end{minipage}
\begin{minipage}{0.4\hsize}
\includegraphics[width=\columnwidth]{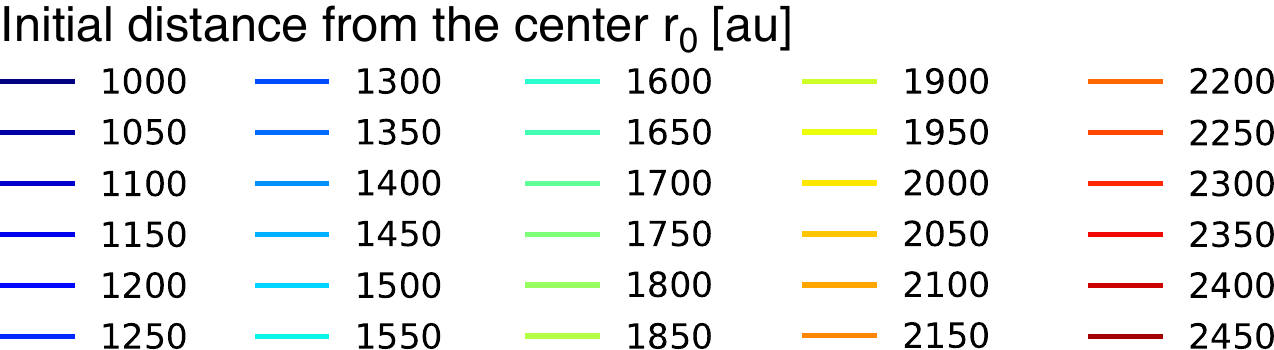}
\end{minipage}
\end{tabular}
\caption{ 
Distance of dust particles from the center against the elapsed time after the cloud collapse begins.
Each panel shows the distance of dust with $\theta_0 = 0, 15, 30, 45, 60, 75, 90^\circ$ (from left top to right bottom).
The dust grain size is $a_{\rm d} = 0.01 \ {\rm \mu m}$.
The initial distances from the center in units of au for each particle are indicated by color, as given on the bottom right.  
The disk size against the elapsed time is also plotted in gray in each panel. 
}
\label{fig:tr2d001mum}
\end{figure*}

\begin{figure*}
\begin{tabular}{cc}
\begin{minipage}{0.4\hsize}
\includegraphics[width=\columnwidth]{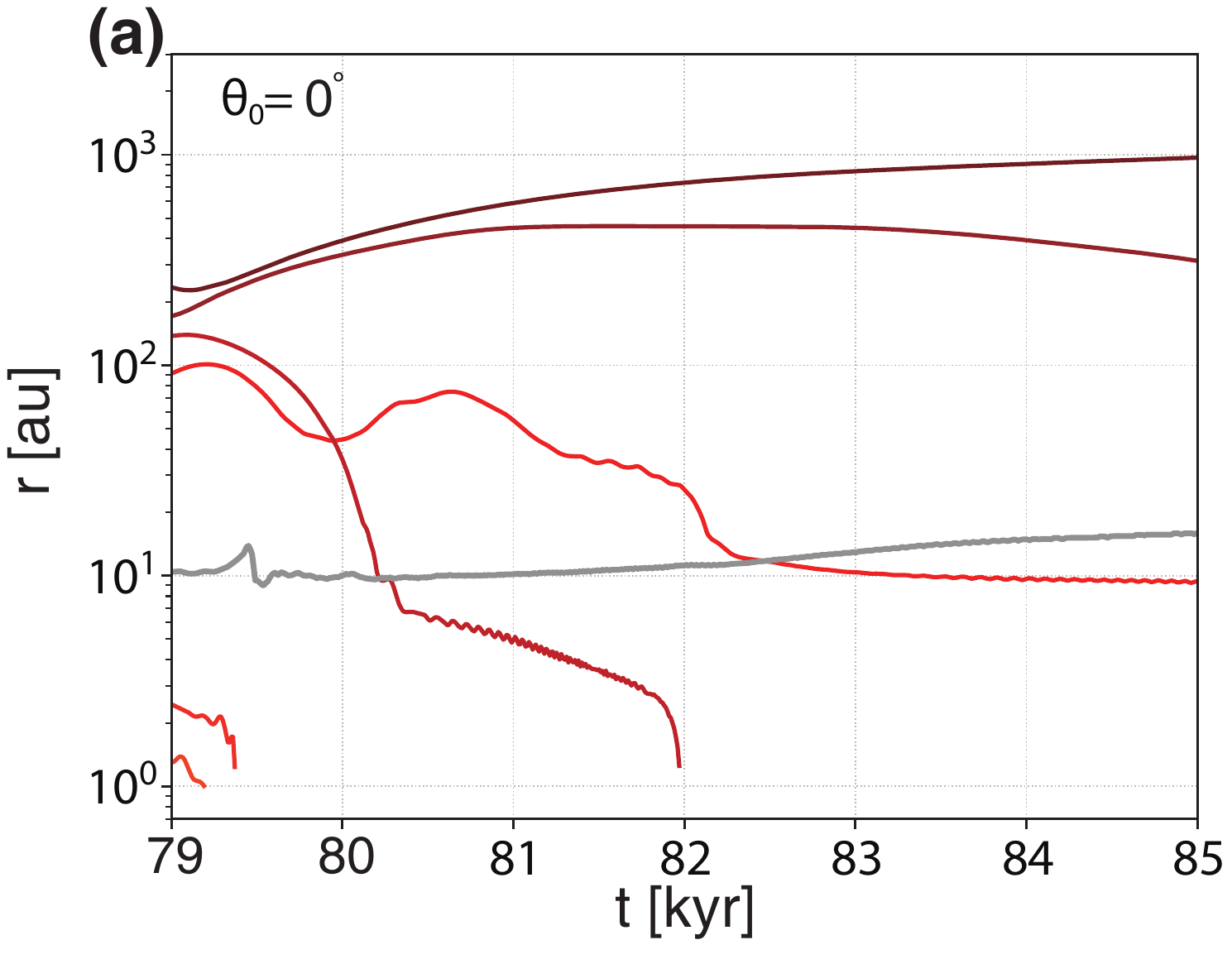}
\end{minipage}
\begin{minipage}{0.4\hsize}
\includegraphics[width=\columnwidth]{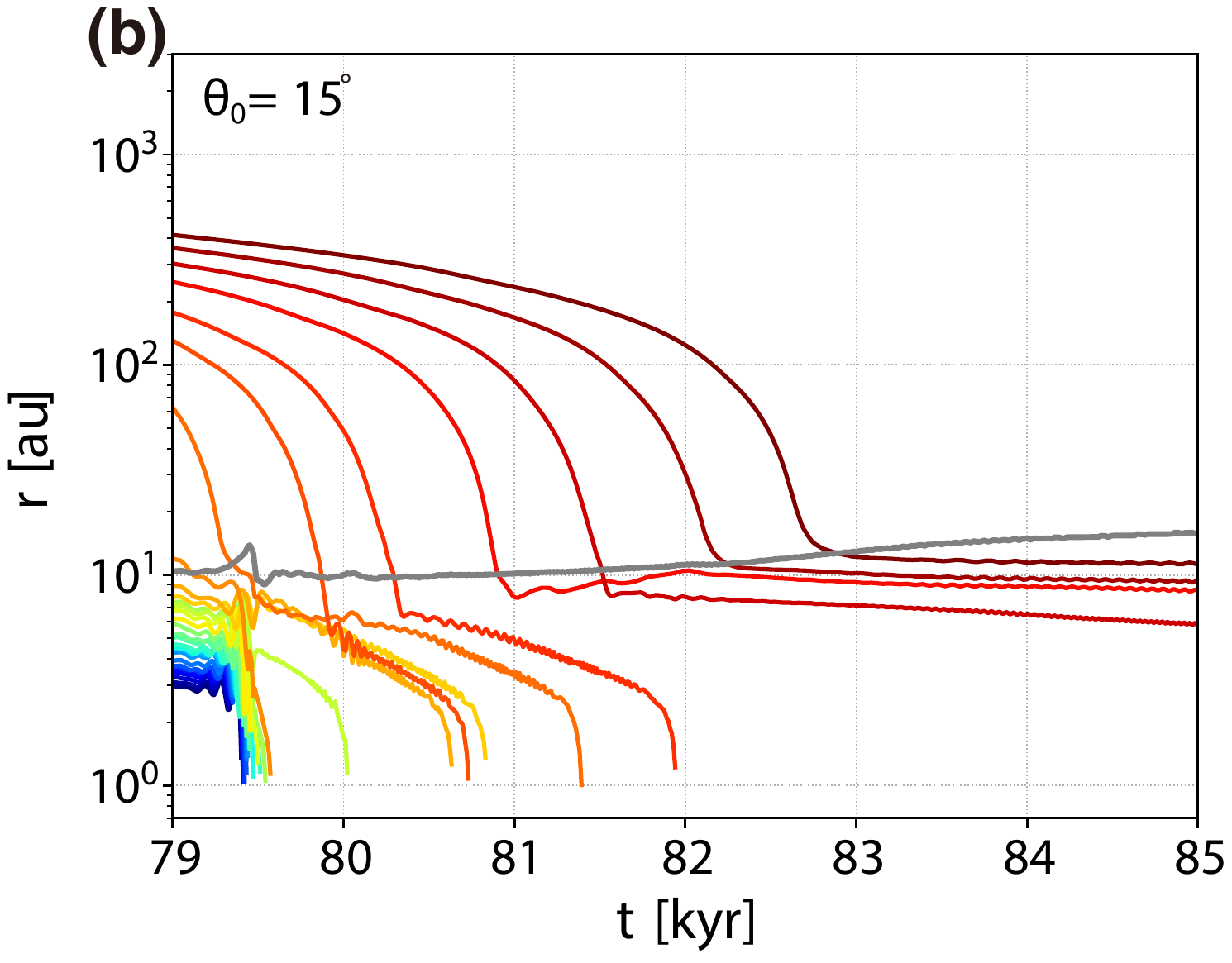}
\end{minipage}
\\
\\
\begin{minipage}{0.4\hsize}
\includegraphics[width=\columnwidth]{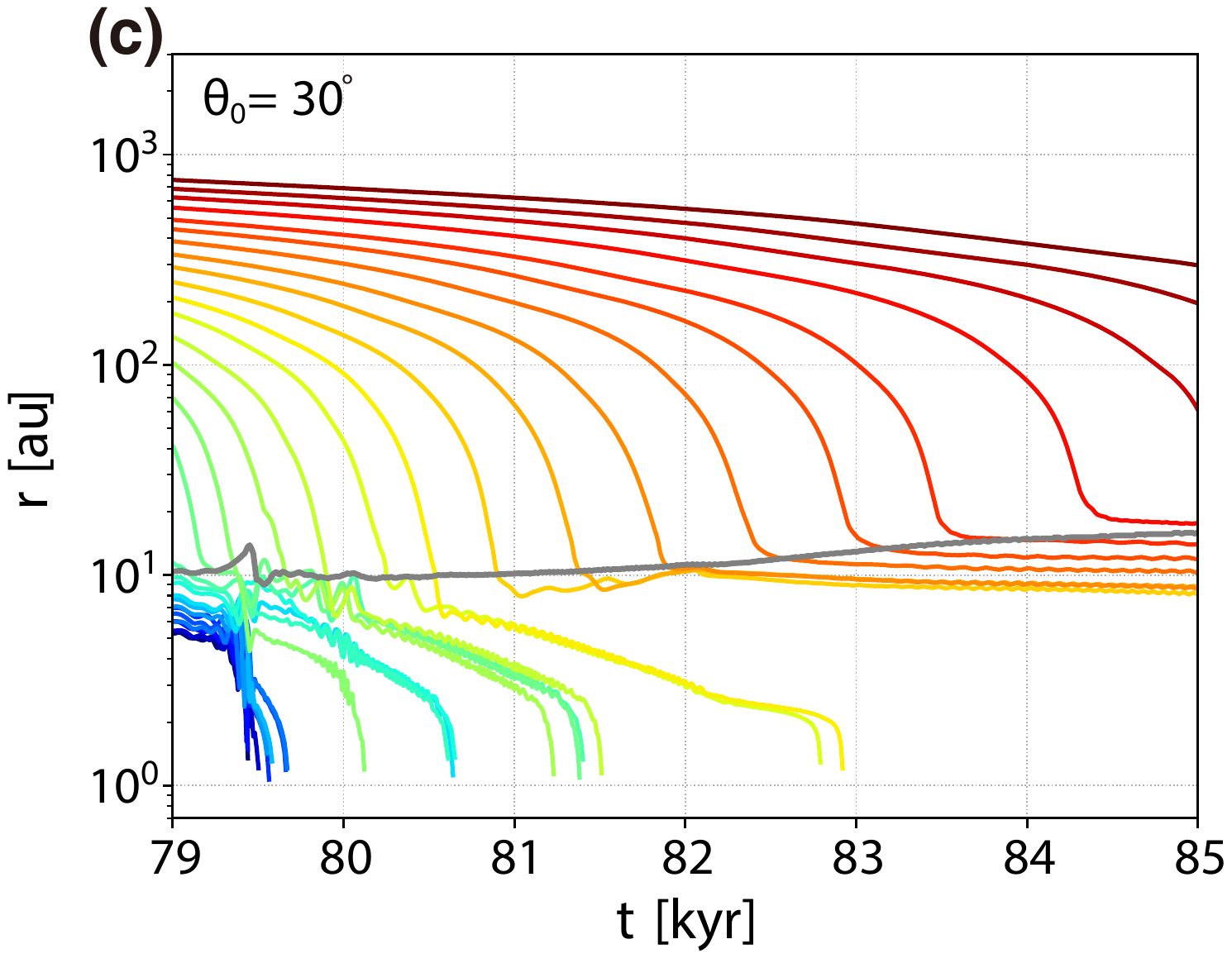}
\end{minipage}
\begin{minipage}{0.4\hsize}
\includegraphics[width=\columnwidth]{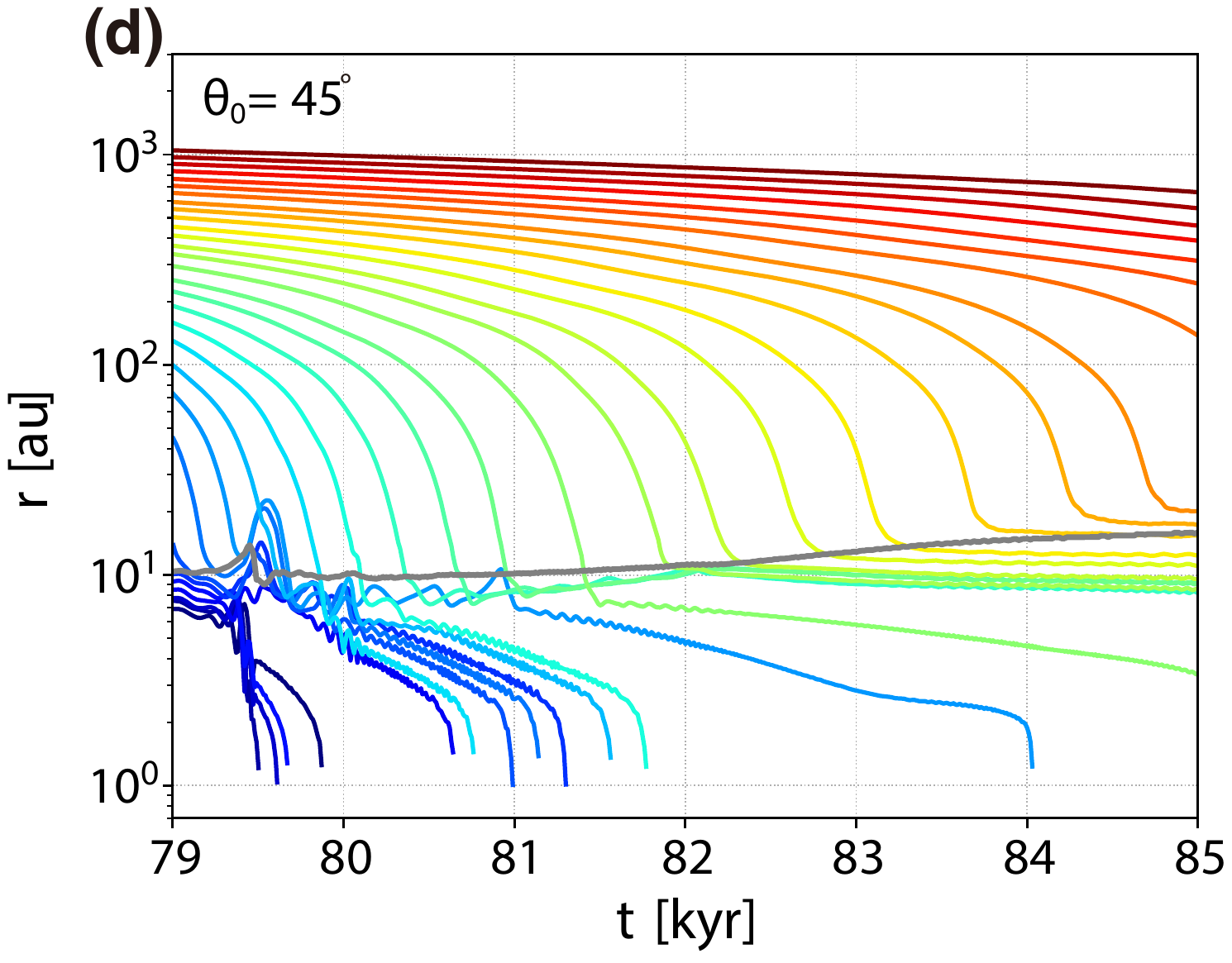}
\end{minipage}
\\
\\
\begin{minipage}{0.4\hsize}
\includegraphics[width=\columnwidth]{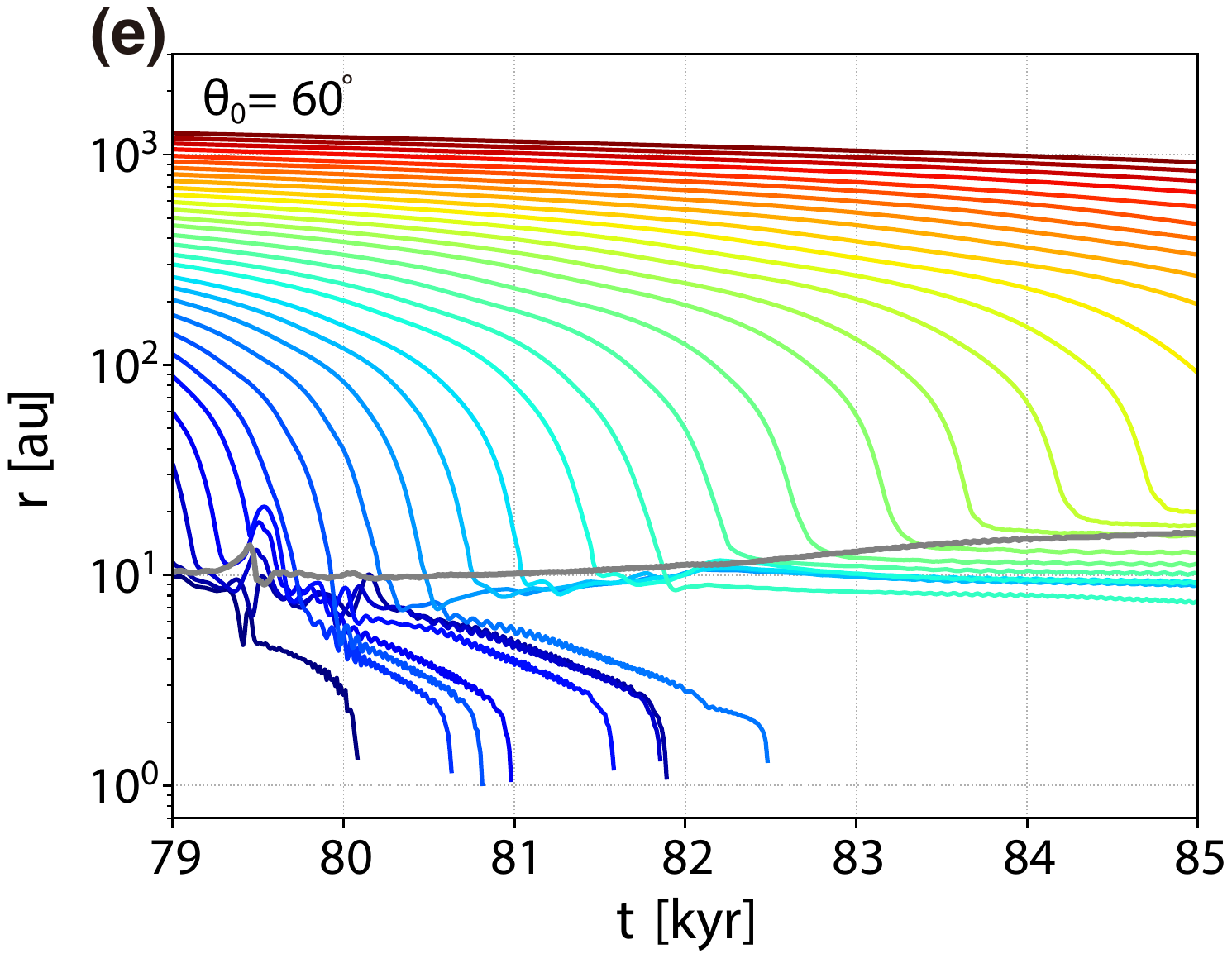}
\end{minipage}
\begin{minipage}{0.4\hsize}
\includegraphics[width=\columnwidth]{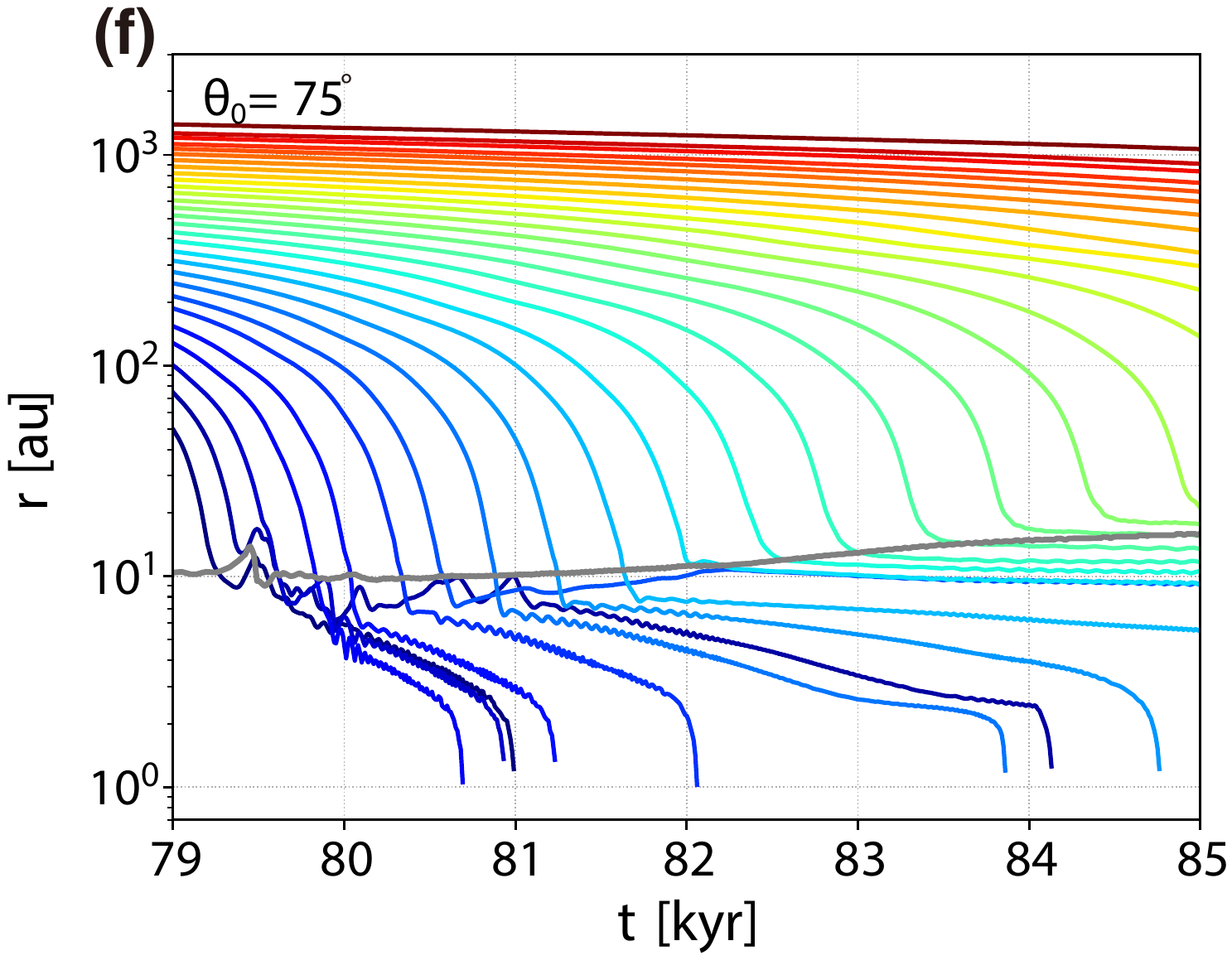}
\end{minipage}
\\
\\
\begin{minipage}{0.4\hsize}
\includegraphics[width=\columnwidth]{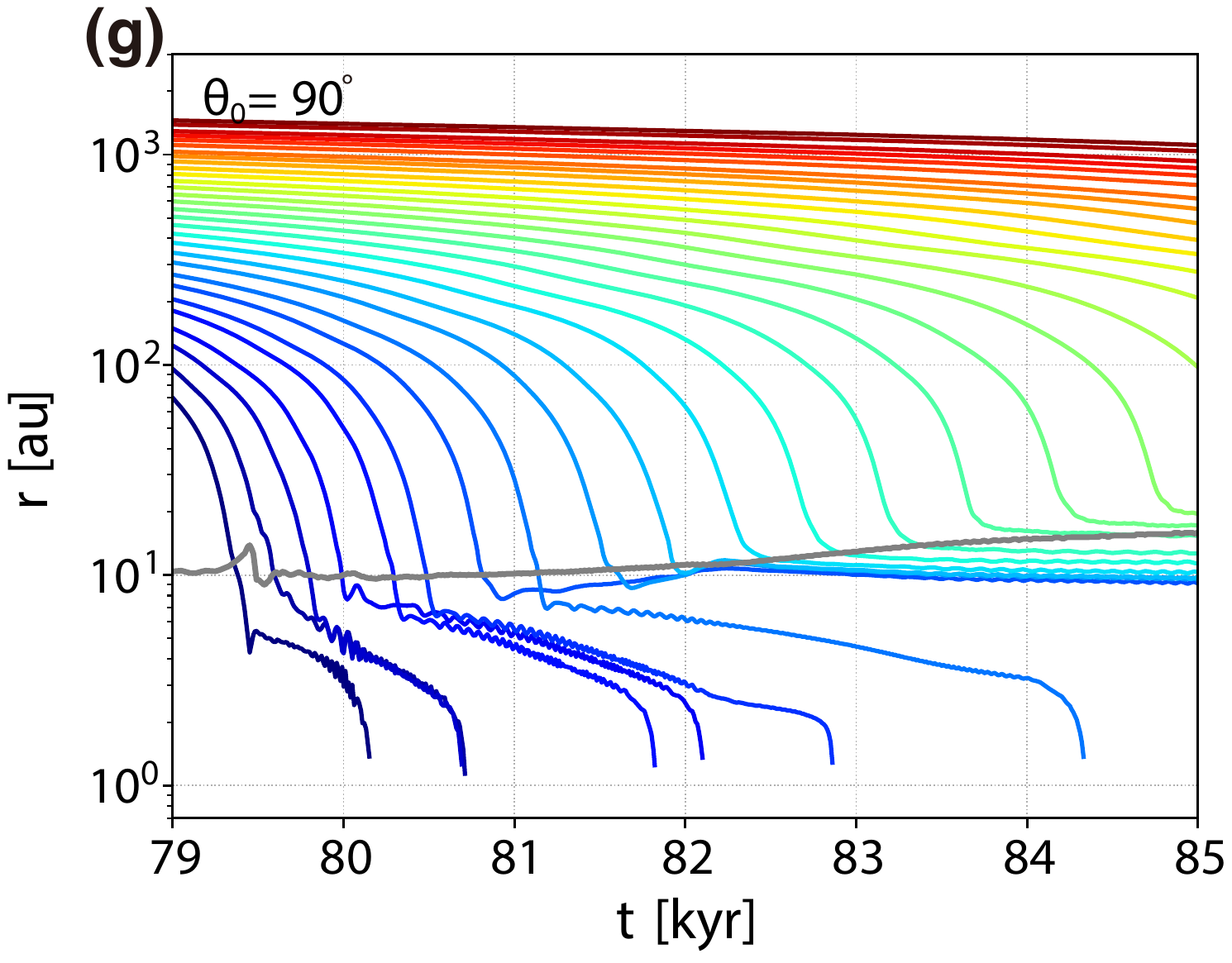}
\end{minipage}
\begin{minipage}{0.4\hsize}
\includegraphics[width=\columnwidth]{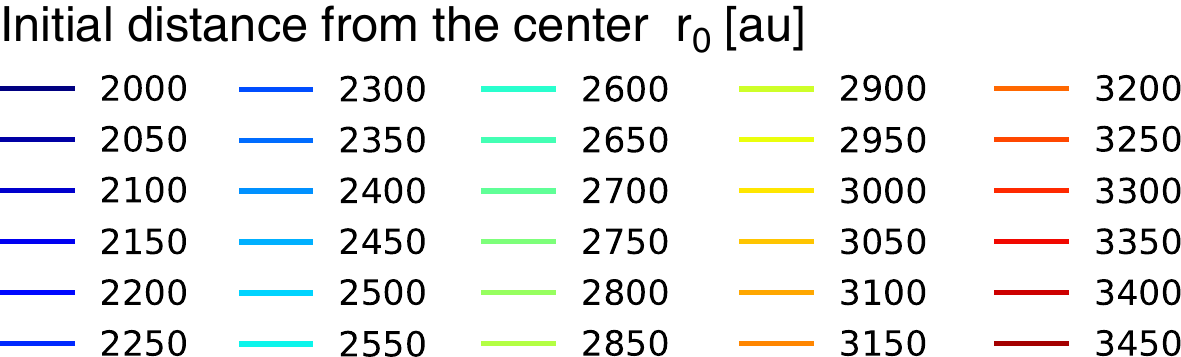}
\end{minipage}
\end{tabular}
\caption{
As Fig.~\ref{fig:tr2d001mum}, for  $a_{\rm d} = 1000\,{\rm \mu m}$.
}
\label{fig:tr2d1000mum}
\end{figure*}

Figs.~\ref{fig:tr2d001mum} and \ref{fig:tr2d1000mum} show the time evolution of the distance from the central protostar  (or sink), $ r=\sqrt{x^2+y^2+z^2}$,  for dust grains with sizes of $a_{\rm d}=0.01\,{\rm \mu m}$  (Fig.~\ref{fig:tr2d001mum}) and $1000\,{\rm \mu m}$ (Fig.~\ref{fig:tr2d1000mum}), respectively. 
In each panel, the initial zenith angle of the dust particles ($\theta_0 = 0, 15, 30, 45, 60, 75$, and $90^\circ$) is fixed.

Firstly, we describe the evolution of dust particles with a size of  $a_{\rm d} = 0.01\,{\rm \mu m}$ (Fig.~\ref{fig:tr2d001mum}). 
In Fig.~\ref{fig:tr2d001mum}, we plot the evolution of dust particles having different initial radial positions in the range $r_0 = 1000$--$2450\, {\rm au}$.  
All the dust particles that initially have  $\theta = 0^\circ$ reach the sink by $t = 79.3\,{\rm kyr}$ (Fig.~\ref{fig:tr2d001mum}{\it a}). 
These particles fall toward the center (or sink) due to the gravity of the central protostar, even though the gas pressure gradient somewhat retards their infall. 
In addition, the dust particles plotted in Fig.~\ref{fig:tr2d001mum}{\it a}  are not swept up by the outflow because they have fallen onto the  sink before the outflow appears. 

The distance evolution for dust particles with $\theta_0 = 15^\circ$ clearly exhibits two trends depending on the initial radius, as seen in  Fig.~\ref{fig:tr2d001mum}{\it b}. 
Dust particles with an initial radius smaller than $r_0 \lesssim 1900$\,au fall onto the disk (gray line) within $t<79.5$\,kyr. 
They then orbit within the disk and finally fall onto the sink by $t\simeq80$\,kyr.
Note that  the gray line in each panel corresponds to the radius of the rotationally supported disk. The identification prescription of the disk is described in Paper I and \S\ref{subsec:dustcouple}. 
Note also that we confirmed that the particles are located within the disk when the distance of each particle is smaller than the disk radius (or gray line). 
On the other hand, all the dust particles having  an initial radius larger than $r_0 \gtrsim1900$\,au increase their distances from the center with time, indicating that they are outflowing from the center with the gas (see Fig.~\ref{fig:outflow}).

For dust particles with $\theta_0 = 30^\circ$ (Fig.~\ref{fig:tr2d001mum}{\it c}), the distance for the  particles with an initial radius larger than $r_0 \gtrsim 1500$\,au begins to increase after initially decreasing, indicating that  these particles  are trapped and entrained by the gas outflow while falling toward the center.
Thus, they fall toward the center at an early stage and outflow from the center at a later stage.
The other particles having an initial radius smaller than $r_0<1500$\,au orbit within the disk and fall onto the protostar, except for one particle plotted in Fig.~\ref{fig:tr2d001mum}{\it c} with an initial radius $r_0=1650$\,au, which survives without falling to the center. 
This surviving particle orbits  within the disk while maintaining a distance of $\sim10$\,au after $t>79$\,kyr. 
In addition, this survivor does not immediately fall to the equatorial plane, but rotates 10 times in the range $0<z<3$\,au within  the hydrostatic equilibrium region of the gas disk before it reaches the equatorial plane of the disk (for details, see below).

A similar trend can be confirmed for the dust particles with $\theta_0 = 45^\circ$ (Fig.~\ref{fig:tr2d001mum}{\it d}). 
The particles initially placed far from the center eventually outflow from the center, while those initially placed near the center fall onto the sink after they enter into the disk region.  
On the other hand, we confirm that particles  at $r_0 = 1850$ and $1900\, {\rm au}$ continue to fall into the center after they are momently rolled up by the outflow.  
Even when the particles are initially placed at almost the same region, their final fates are somewhat different. 
The outflow launching region varies with time, and the timing of the dust particles reaching the disk surface is influenced by whether they are swept up by the outflow.

For the dust particles with $\theta_0 \ge60^\circ$ (Fig.~\ref{fig:tr2d001mum}{\it e}--{\it g}), we see no rapid increase in the distance from the center, indicating that the particles plotted in these panels are not captured by the outflow. 
After the dust particles initially located near the center ($r_0\lesssim1500$\,au) reach the disk, they orbit in the disk while gradually falling to the center,  as seen in Fig.~\ref{fig:tr2d001mum}{\it e}--{\it g}. 
On the other hand, the particles initially located far from the center ($r_0\gtrsim1800$\,au) orbit within the disk without falling onto the protostar by the end of the simulation.
After these particles reach the disk, they orbit in the disk while maintaining a distance of $\sim10$--$20$\,au.
The distances from the center for these surviving particles are comparable to and slightly smaller than the disk radius.
Thus, these particles orbit around the outer disk region by the end of the simulation.
Although the protostar and disk gradually evolve,  these particles continue to orbit in the disk maintaining almost the same radii. 
As a result, among the dust particles with $\theta_0 \gtrsim 60^\circ$, there are many survivors orbiting in the disk outer region for a long time.  

Next, we describe the behavior of the dust grains with a size of  $a_{\rm d}=1000\,{\rm \mu m}$ shown  in Fig.~\ref{fig:tr2d1000mum}.
Note that the initial distances of the particles from the center are different from Fig.~\ref{fig:tr2d001mum}. 
We plotted only the dust particles initially placed in the range $r_0=2000$--$3450$\,au, because most of the dust particles  having initially small radii ($r_0 \lesssim 2000$\,au) fell directly onto the sink without passing through the circumstellar disk (Paper I).
Similar to the case for $a_{\rm d}= 0.01\, {\rm \mu m}$  (Fig.~\ref{fig:tr2d001mum}), the dust particles have three major paths, falling onto the sink, being trapping by the rotationally supported disk, and being ejected by the outflow.

Most of the dust particles plotted in Fig.~\ref{fig:tr2d1000mum} accrete onto the disk. 
Ejection by the outflow can be seen for only  a few particles with $\theta_0=0^\circ$ (Fig.~\ref{fig:tr2d1000mum}{\it a}). 
In Fig.~\ref{fig:tr2d1000mum}{\it a}, we can confirm that the distance from the center continues to increase with time for one particle ($r_0 = 3450$\,au), which is outflowing from the center. 
The particles with $r_0=3400$,  3350, and 3300\,au are entrained by the outflow and their distances initially increase
but then decrease, indicating that they fall to the center. 
For $\theta_0 \ne0$, the dust particles orbit within the disk after they reach the disk.
In Fig.~\ref{fig:tr2d1000mum}{\it b}-{\it g}, the dust particles initially placed in the range of  $r_0 \lesssim 2500$\,au finally falls onto the sink. 
On the other hand, the particles initially placed in the range $r_0 \gtrsim 2500$\,au maintain relatively constant distances from the center by the end of the simulation after they enter the disk.
The rotationally supported disk has a size of $\sim10$--$20$\,au
and these particles move with almost the same orbit of $r\sim10$--$20$\,au, which is the outer disk region or near the outer edge of the disk.

Comparing Fig.~\ref{fig:tr2d001mum} and Fig.~\ref{fig:tr2d1000mum} shows that the fraction of particles falling onto the disk is larger for dust particles with $a_{\rm d}=1000\,{\rm \mu m}$ than for those with $a_{\rm d}= 0.01\, {\rm \mu m}$. 
The dust particles with a size of $1000\,{\rm \mu m}$ are initially more decoupled from the gas than those with a size of $0.01 \rm {\rm \mu m}$, as described in Paper I. 
Thus, large dust particles tend to fall onto the central region where the disk forms, partially ignoring the gas drag in the infalling envelope.  
For dust with $a_{\rm d}=1000\,{\rm \mu m}$  (Fig.~\ref{fig:tr2d1000mum}), the dust particles orbiting in the circumstellar disk are initially distributed in the range $\theta_0=15$--$90^\circ$, where many particles maintain their distances from the center in the range $10$--$20$\,au. 
Note that one particle ($r_0=3300$\,au) in Fig.~\ref{fig:tr2d1000mum}{\it a} survives in the disk without falling onto the sink for $\theta_0=0^\circ$.
On the other hand, a large fraction of the dust particles with $a_{\rm d}= 0.01\, {\rm \mu m}$  (Fig.~\ref{fig:tr2d001mum}) initially distributed in the range $\theta_0=15$--$45^\circ$ are ejected by the outflow, after initially moving towards the center. 
Thus, during the simulation, the number of dust particle orbiting within the circumstellar disk is larger for a size of $a_{\rm d}=1000\,{\rm \mu m}$ than for  a size of $a_{\rm d}= 0.01\, {\rm \mu m}$. 

In the above, we only showed the behavior for dust particles with sizes of $a_{\rm d}=0.01$ and 1000\,$\mu$m (i.e., the smallest and largest dust particles considered in this study and Paper I). 
In the following, we describe the dust motion for other particle sizes ($a_{\rm d}$ =0.1, 1, 10, 100\,$\mu$m). 
The behavior of the particles with $a_{\rm d}$=0.1, 1, 10\,$\mu$m is the same as for $a_{\rm d}=0.01\,\mu$m.
Fig.~14 of Paper I showed that the dust abundances of the envelope, protostar, disk, and outflow are the same (or do not change over time) for dust particles in the size range  $a_{\rm d}=0.01$--$10\,\mu$m. 
The Stokes numbers for these particles are much less than unity over the simulation, indicating that these particles are well coupled with the gas at all times\footnote{
The definition of the Stokes number in Paper I is slightly different from that in the present paper (for details, see \S2.3.1 of Paper I and \S\ref{subsec:dustcouple}).
}. 
Therefore, there is no significant difference in the behavior of these particles.  
On the other hand, the abundance of particles with $a_{\rm d}=100\,\mu$m is different from that of particles with $a_{\rm d}\le10\,\mu$m, but is almost the same as for the case of $a_{\rm d}=1000\,\mu$m, because the Stokes number sometimes exceeds unity for these particles (Fig.~17 of Paper I). 
For this reason,  we consider particles with sizes of 0.01 and 1000 $\,\mu$m to illustrate the typical motion of dust grains with different sizes.  

\subsubsection{Dust motion within the disk} 
As described in \S\ref{subsub:envelope}, the dust particles enter the disk when they are not ejected by the outflow.
Within the disk, some particles survive without falling onto the disk.
Figs.~\ref{fig:tr2d001mum} and \ref{fig:tr2d1000mum} show that  there is a wide  range of initial zenith angle $\theta_0=15$--$90^\circ$ for which the dust particles continue to orbit around $10$--$20$\,au.
These `non-drifting' dust particles may provide a solution to the radial drift barrier problem, which is one of the most important issues in theoretical planet formation scenarios.
In the following, we focus on the trajectories of such particles.

\begin{figure}
    \centering
    \includegraphics[width=\linewidth]{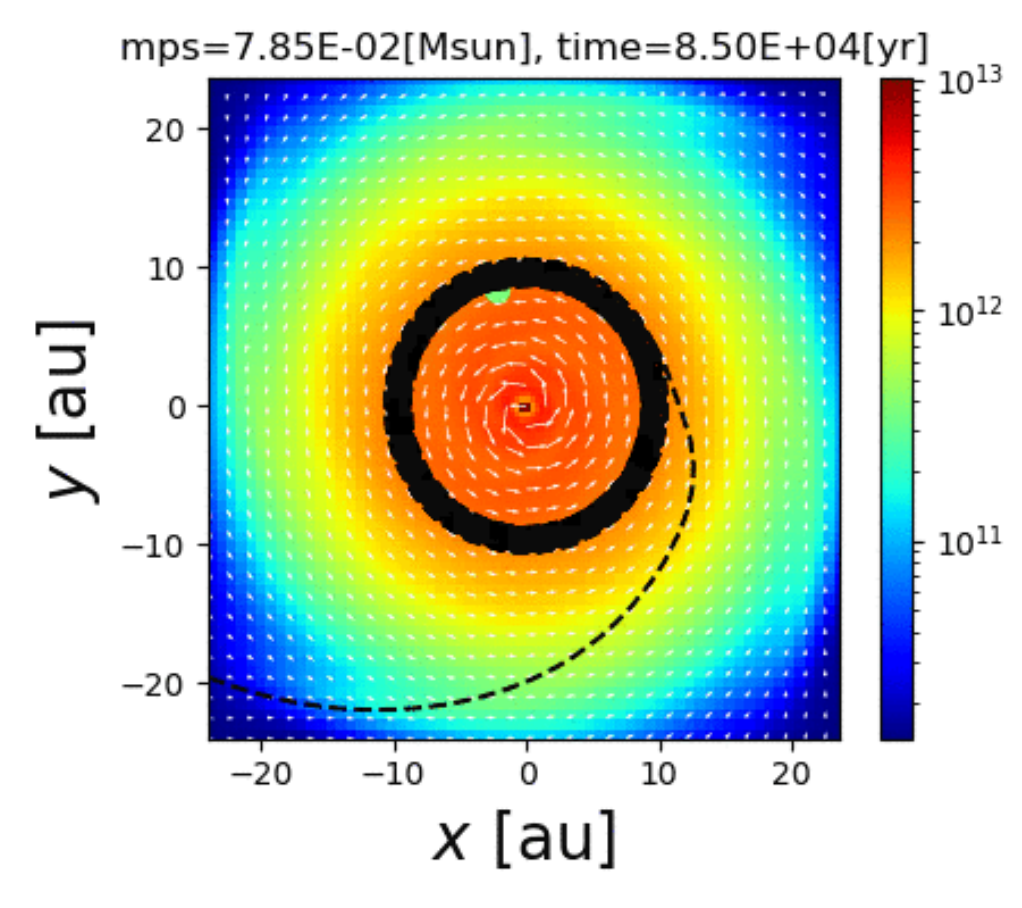}
    \caption{
Trajectory of a dust particle (black broken line) superimposed on  gas density (color) and velocity (arrows) distributions on the $z = 0$ plane. 
The particle is initially placed at ($r_0$, $\theta_0$, $\phi_0$)= (1500\,au, $90^\circ$, $0^\circ$) and has a size of  $a_{\rm d}=0.01\,{\rm \mu m}$. 
The protostellar mass and elapsed time are given above the plot. }
    \label{fig:l13_001mum_th90_r1500}
\end{figure}

Fig.~\ref{fig:l13_001mum_th90_r1500} shows the trajectory of a  dust particle with a size of $a_{\rm d}=0.01\,{\rm \mu m}$, initially placed at ($r_0$, $\theta_0$, $\phi_0$)= (1500\,au, $90^\circ$, $0^\circ$). 
Note that this particle moves only on the $z=0$ plane because it is initially placed on the equatorial plane. 
After the dust particle reaches the rotationally supported disk, it orbits 30 times around the center, keeping a distance of $\sim10$\,au from the center. 
As the figure shows, the particle does not fall to the center by the end of the calculation. 
Although we only plot the trajectory of a single particle in Fig.~\ref{fig:l13_001mum_th90_r1500}, all particles having $\theta_0 = 90^\circ$ continue to orbit around the protostar without falling onto the sink after they reach the rotationally supported disk.
Interestingly, such particles maintain an orbital radius of $\sim10$--$20$\,au during the calculation, as seen in Figs.~\ref{fig:tr2d001mum} and \ref{fig:tr2d1000mum}.

\begin{figure}
    \centering
    \includegraphics[width=\linewidth]{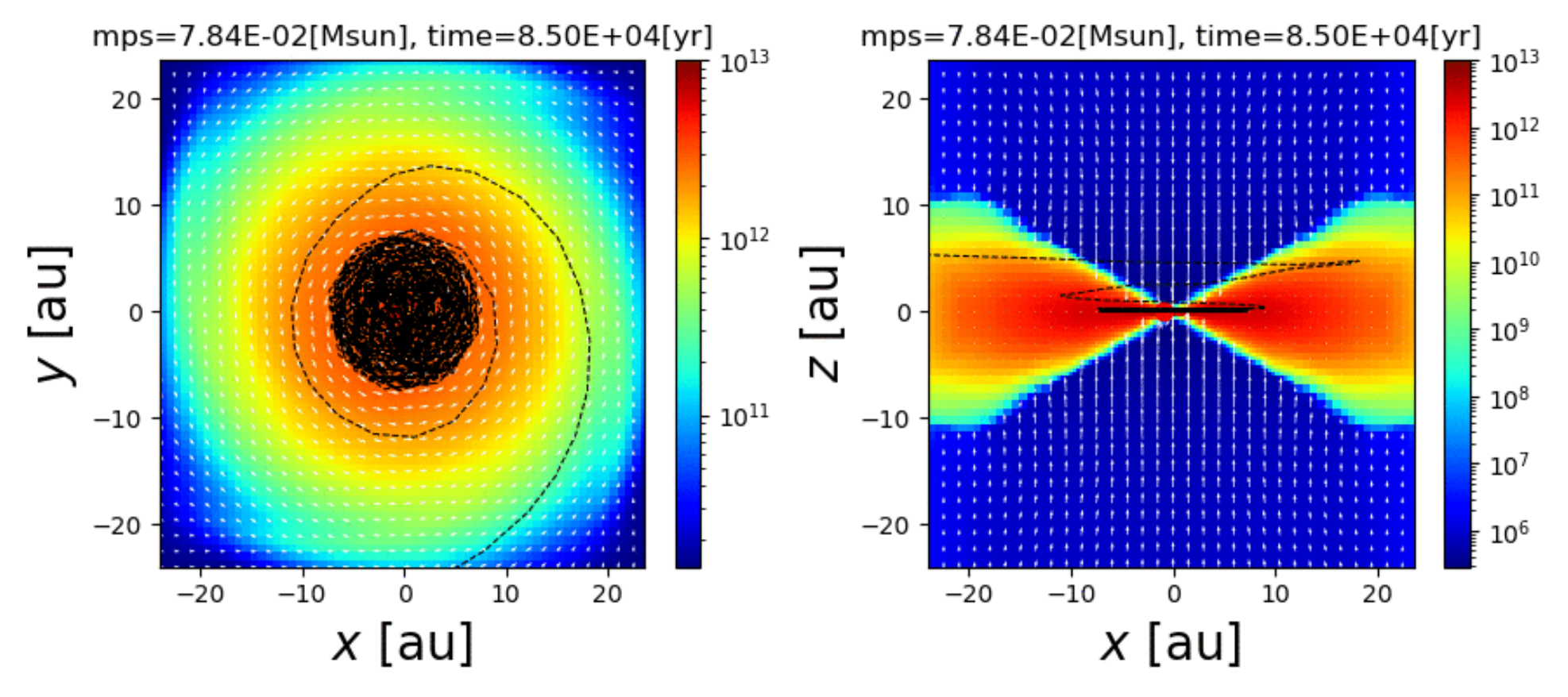}
    \caption{
Trajectory of a dust particle (black broken line) superimposed on  gas density (color) and velocity (arrows) distributions on the $z = 0$ (left) and  $y=0$ (right) planes. 
The particle is initially placed at ($r_0$, $\theta_0$, $\phi_0$)= (1900\,au, $45^\circ$, 0$^\circ$) and has the size of  $a_{\rm d}=0.01\,{\rm \mu m}$. 
The protostellar mass and elapsed time are given above the plot. 
}
    \label{fig:l13_001mum_th45_r1900}
\end{figure}

Next, we describe the trajectory of a dust particle that ultimately falls onto the sink. 
Fig.~\ref{fig:l13_001mum_th45_r1900} shows the trajectory of a particle with $a_{\rm d} = 0.01\, {\rm \mu m}$, initially placed at ($r_0$, $\theta_0$, $\phi_0$)= (1900\,au, $45^\circ$, 0$^\circ$). 
As seen in the right panel of Fig.~\ref{fig:l13_001mum_th45_r1900}, the particle gradually approaches the equatorial plane while going around the center before it enters the disk region. 
This particle enters the disk from the vertical direction, whereas the particle shown in Fig.~\ref{fig:l13_001mum_th90_r1500}, initially placed on the equatorial plane, enters the disk from the disk outer edge on the equatorial plane. 
As seen in the left panel of Fig.~\ref{fig:l13_001mum_th45_r1900}, the particle gradually falls toward the center after it reaches the equatorial plane. 
The particle orbits around the center for about 4000\,yr, slowly  moving in a radial direction, while the distance from the center continues to decrease. 
The particle finally falls onto the sink (or protostar).

Although we only show two typical cases above, the same trend can be seen for other particles.  
When the dust particles are initially placed near the center (small $r_0$), they spiral to the sink after reaching the equatorial plane of the disk. 
The orbital radii for these particles are $r\lesssim 10$\,au when they reach the equatorial plane.  
Dust particles initially placed far from the center (large $r_0$) settle around $r\gtrsim10$\,au in the disk by the end of the simulation. 
Thus, it can be concluded that dust particles will survive without falling onto the sink when they have an orbital radius of $\gtrsim 10$\,au.
These particles tend to reach the outer disk region ($\simeq 10$--$20$\,au), passing through the disk outer edge, as shown in Fig.~\ref{fig:l13_001mum_th90_r1500}. 

\subsection{Dust coupling in the disk}
\label{subsec:dustcouple}
\begin{figure}
    \centering
    \includegraphics[width=\linewidth]{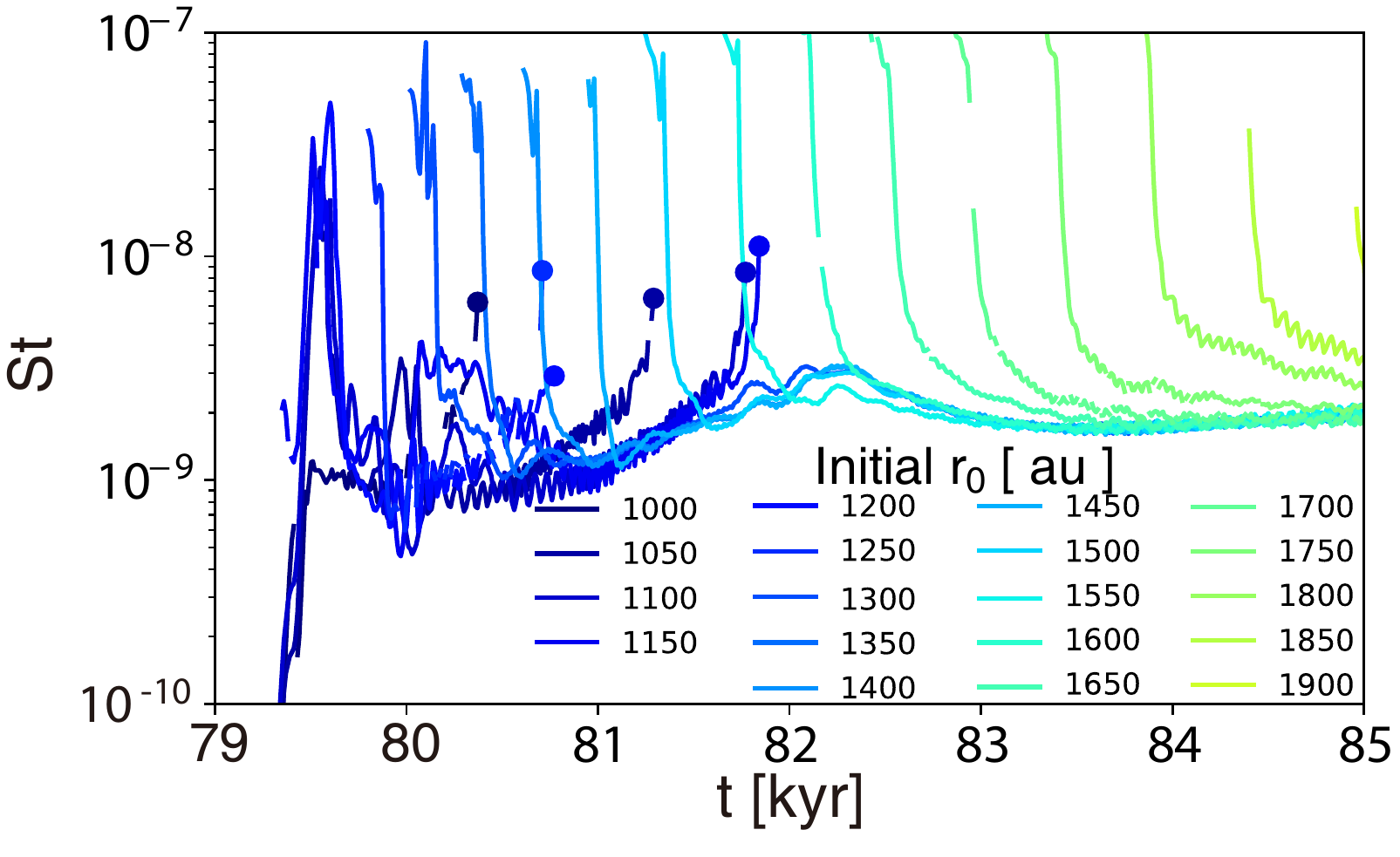}
    \caption{
Time evolution of the Stokes number for dust particles with $a_{\rm d} = 0.01\,{\rm \mu m}$. 
St is defined in equation (\ref{eq:sttk}). 
Lines are plotted only when the particles are moving in the disk region.
The color indicates the initial radius $r_0$ and is  the same as that  in Fig.~\ref{fig:tr2d001mum}.
Filled circles represent particles that fall onto the sink.
}
\label{fig:tSt_th90_r1000to2500_001mum_disk}
\end{figure}

\begin{figure}
    \centering
    \includegraphics[width=\linewidth]{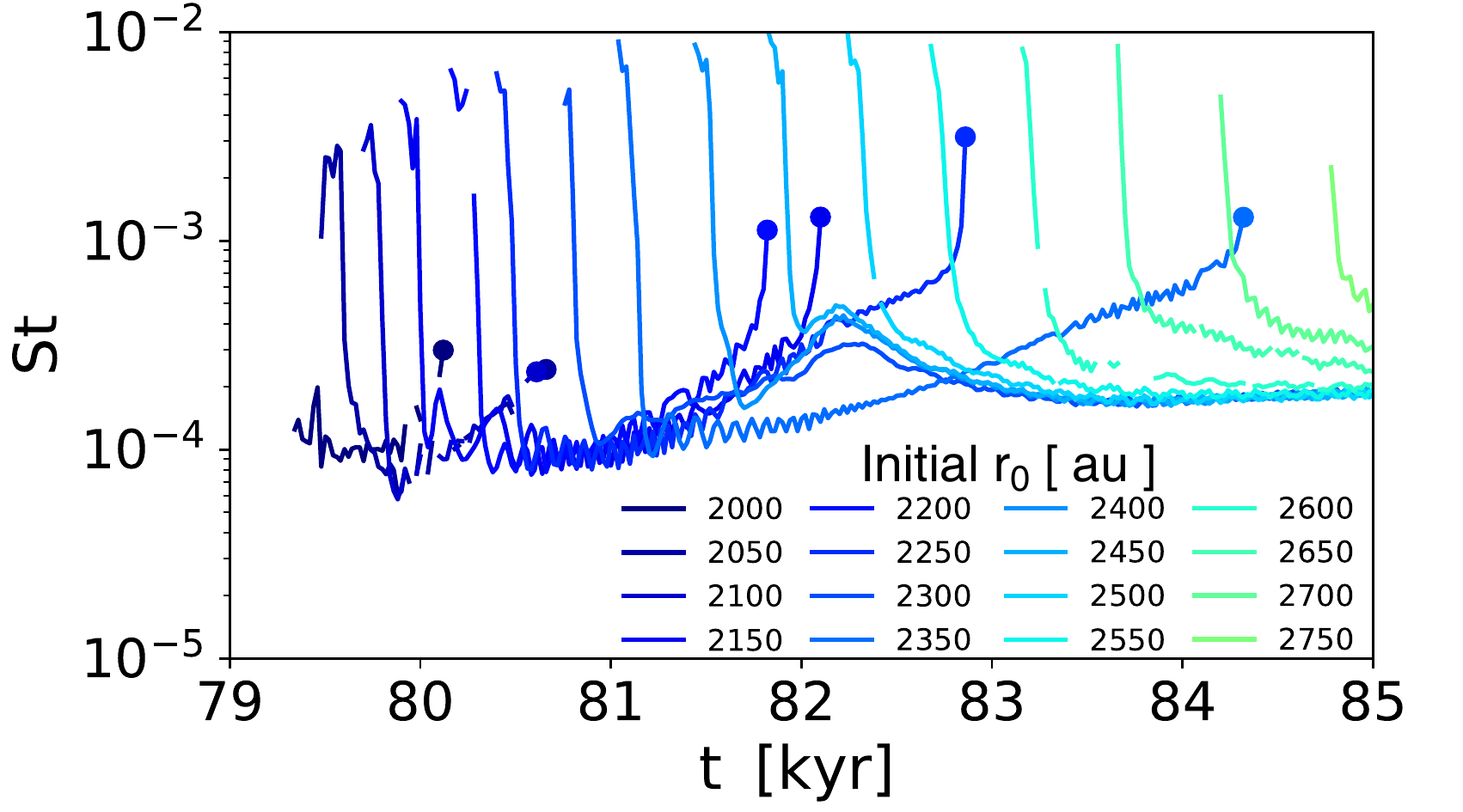}
    \caption{
As Fig.~\ref{fig:tSt_th90_r1000to2500_001mum_disk}, for particles with $a_{\rm d} = 1000\,{\rm \mu m}$. 
The color indicates the initial radius $r_0$ and is  the same as that  in Fig.~\ref{fig:tr2d1000mum}.}
    \label{fig:tSt_th90_r2000to3500_1000mum_disk}
\end{figure}

In section \S \ref{subsec:dustr}, we presented dust trajectories that are not expected based on classical planet formation scenarios. 
In this subsection, we investigate whether the dust particles that move in the disk are coupled with the gas.

In \S\ref{subsec:dustr}, we showed curious behavior of dust grains, in which some dust grains continue to orbit at the same orbital radius without moving toward the center. 
If the dust grains are strongly coupled with the gas in the disk, the dust motion should coincide  with the gas motion. 
If this is the case,  the gas motion can explain why some dust particles maintain an orbit around  $r \simeq 10$--$20$\,au.
The gas density is higher in the disk than in the envelope
and thus, assuming same-sized dust particles, the coupling between the dust particles and  gas is stronger in the disk than in the envelope. 
To quantitatively evaluate how strongly a dust particle is coupled with the gas in the disk, we use the Stokes number St, defined as 
\begin{equation}
{\rm St} = \frac{t_{\rm s}} {t_{\rm K}},
\label{eq:sttk}
\end{equation}
where $t_{\rm K}$ is the Keplerian  timescale $\Omega_{\rm kep}^{-1}$ ($=(GM_{\rm ps}/r)^{-1/2}$).

Figs.~\ref{fig:tSt_th90_r1000to2500_001mum_disk} and \ref{fig:tSt_th90_r2000to3500_1000mum_disk} show the time evolution of the Stokes number for dust particles orbiting within  the disk with $a_{\rm d} = 0.01$ and $1000\,{\rm \mu m}$. 
According to Paper I, the disk is defined as the region where the rotational velocity dominates the infall velocity ($v_\phi > 2 |v_r|$) and the gas is rotationally supported  ($v_\phi > 0.6\,v_{\rm K}$, where $v_{\rm K}$ is the Keplerian velocity). 
In the figures, a broken part of the line indicates that the dust particle is temporarily located outside the disk. 
When the line is broken while St decreases, the particle has temporarily moved into the envelope near the disk. 
A filled circle indicates the epoch at which  the dust particle falls onto the sink before the end of the calculation ($t<85$\,kyr). 
A sudden increase in St is caused by a sudden decrease in the radius of a particle in the disk.

Fig.~\ref{fig:tSt_th90_r1000to2500_001mum_disk} shows that dust particles entering the disk from the envelope have  St $\ll 1$. 
In particular, the Stokes number for dust particles orbiting at  $r \simeq 10$--$20$\,au is  ${\rm St}\sim 10^{-9}$, indicating that they are  strongly coupled with the gas. 
Fig.~\ref{fig:tSt_th90_r2000to3500_1000mum_disk} plots the time evolution of St for dust with $a_{\rm d} = 1000\, {\rm \mu m}$, which is the maximum dust  size prepared in this study. 
When the gas has  the same physical quantities, larger dust particles tend to be more easily decoupled  from the gas due to their longer stopping time.
However, even for dust particles having $a_{\rm d} = 1000\, {\rm \mu m}$,  the Stokes number is  ${\rm St} \sim 10^{-4}$.
Thus, such  large particles are also coupled with the gas in the disk. 

The St for particles with $a_{\rm d} = 1000\, {\rm \mu m}$ is about five orders of magnitude larger than that for dust particles with $a_{\rm d} = 0.01\, {\rm \mu m}$.
This is because the Stokes number is proportional to the dust size as  ${\rm St}\propto a_{\rm d}$. 
Therefore,  we can conclude that the result that some dust particles orbit at the same orbital radius instead of gradually falling toward the central star can be attributed to the gas motion, that is, the  gas motion on the equatorial plane in the disk  prevents the dust particles from accreting onto the central protostar.
Thus,  the dust grains are controlled by gas motion in the disk independent of the dust size in the early star formation stage.

\section{Discussion}
\subsection{Angular momentum transport and  gravitational instability}
As shown in \S \ref{subsec:dustcouple},  a portion of the dust particles keep orbiting around $r\simeq10$--$20$\,au in the disk without falling onto the sink.  
The dust particles eternally orbiting in the disk are strongly coupled with the gas, because  the gas density is sufficiently high and the Stokes number is much smaller than unity. 
Thus, to understand why some of the dust is captured in an orbit, it is important to investigate the physical properties of the gas.
In this subsection, we investigate the characteristics of the gas disk around the central region.

\begin{figure}
    \centering
    \includegraphics[width=\linewidth]{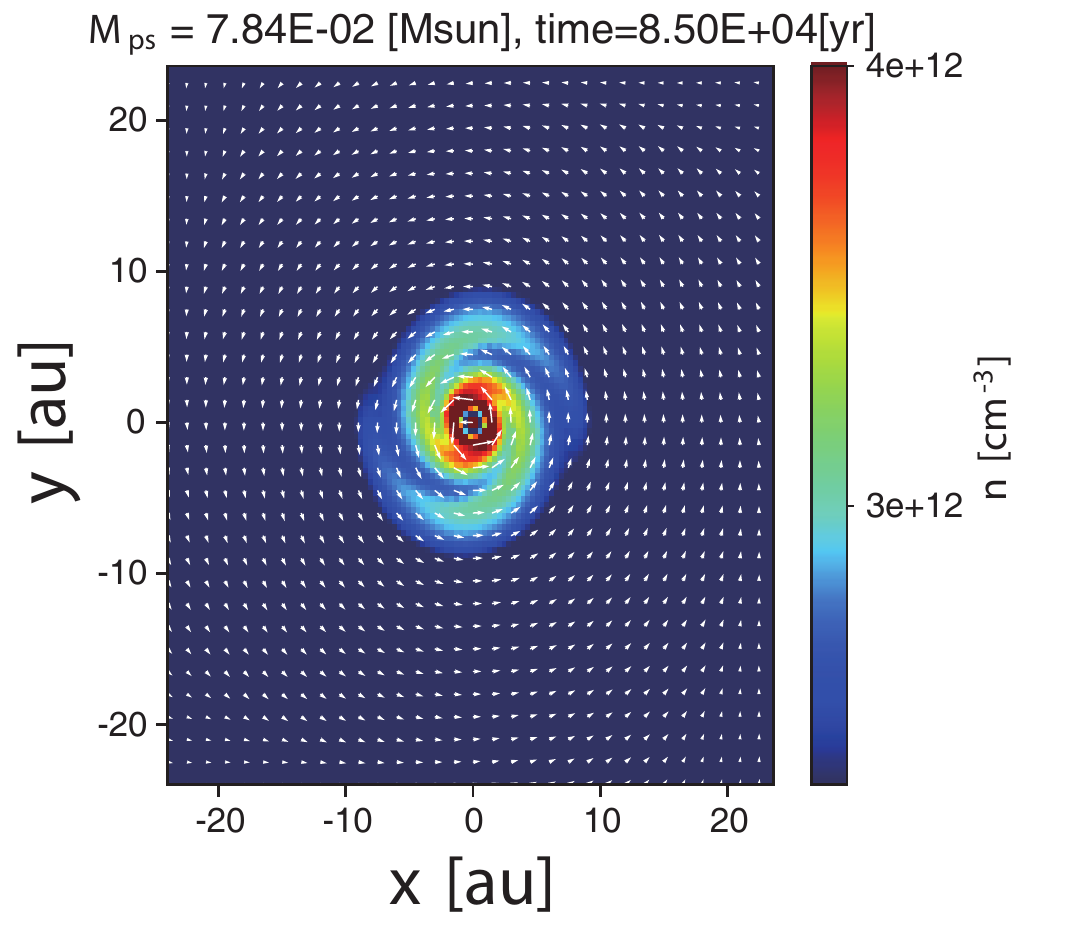}
    \caption{Same as the bottom left panel of Fig.~\ref{fig:gevol}, with a different color (or density) range.}
    \label{fig:l13spiral}
\end{figure}

First, we focus on the angular momentum transport in the disk. 
To extract the structure in the high-density region, we plot the density distribution in the density range  $n>2.5\times10^{12}\,\cc$ in Fig.~\ref{fig:l13spiral}.
The figure clearly shows that the local high-density region has a spiral structure which extends to $\sim10$\,au. 
The elapsed time and spatial scales are the same as in the bottom left panel of Fig.~\ref{fig:gevol}. 
Previous studies have shown  the formation of a spiral structure driven  by gravitational instability \citep{2017ApJ...835L..11T}.
Since the high-density spiral arm has a non-axisymmetric structure, the angular momentum could be transported by the  gravitational torque.

\begin{figure}
    \centering
    \includegraphics[width=\linewidth]{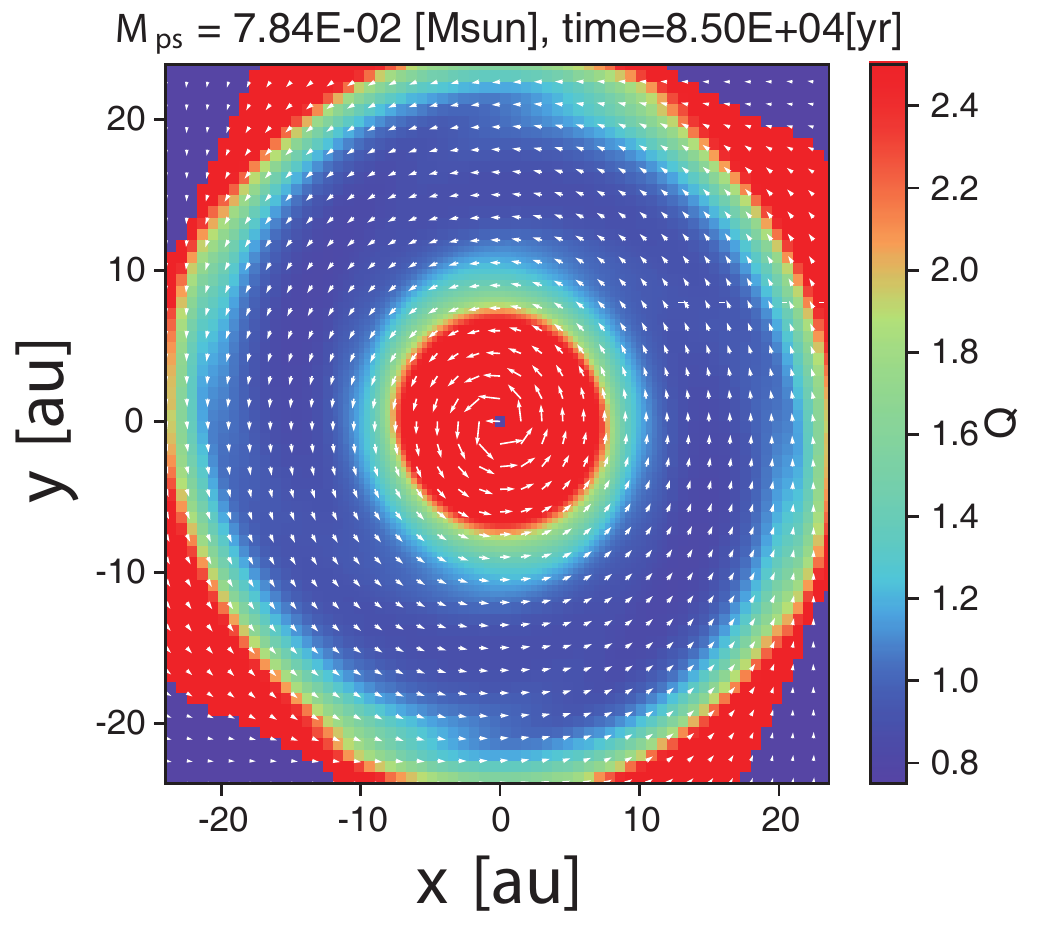}
    \caption{Toomre's $Q$ parameter (color) for the disk plotted on the $z$ = 0 plane, in which the physical quantities are integrated in the vertical direction within the disk and are averaged. }
    \label{fig:l13q}
\end{figure}

To validate the gravitational instability in the disk, the Toomre $Q$ parameter is plotted in Fig.~\ref{fig:l13q}. 
The definition and calculation method of the $Q$ parameter in the disk  are described  in Paper I and \citet{2017ApJ...835L..11T}. 
The figure indicates that $Q < 2$ in the range $7\,{\rm au} \lesssim r \lesssim 20\,{\rm au}$.
Thus, a spiral structure will naturally develop in this region. 
The spiral arms extend to $\sim10$\,au (Fig.~\ref{fig:l13spiral}), while the $Q$ parameter has a local minimum around $r\sim10$--$15$\,au (Fig.~\ref{fig:l13q}).  
The position of the spiral arms is not in complete agreement with the region having a local minimum of $Q$, as previously shown in \citet{2017ApJ...835L..11T}. 
The angular momentum should be transported outward due to gravitational torque caused by the spiral arms inside the $\lesssim10$\,au region of the disk. 
In contrast, the gas distributed in the outer disk region of $r\gtrsim 10$\,au should gain angular momentum due to angular momentum conservation.
Thus, a parcel of gas can stay around $r\simeq10$--$20$\,au without moving inward. 
Therefore, the gas around  $r\gtrsim 10$\,au orbits at almost the same radius  while maintaining its angular momentum. 
As seen in Figs.~\ref{fig:tr2d001mum}--\ref{fig:tSt_th90_r2000to3500_1000mum_disk}, the dust particles in the disk are strongly coupled with the gas.
As a result, the dust particles located around $r\gtrsim 10$\,au receive angular momentum from the gas and survive to orbit in the disk without causing inward radial drift. 

\begin{figure}
    \centering
    \includegraphics[width=\linewidth]{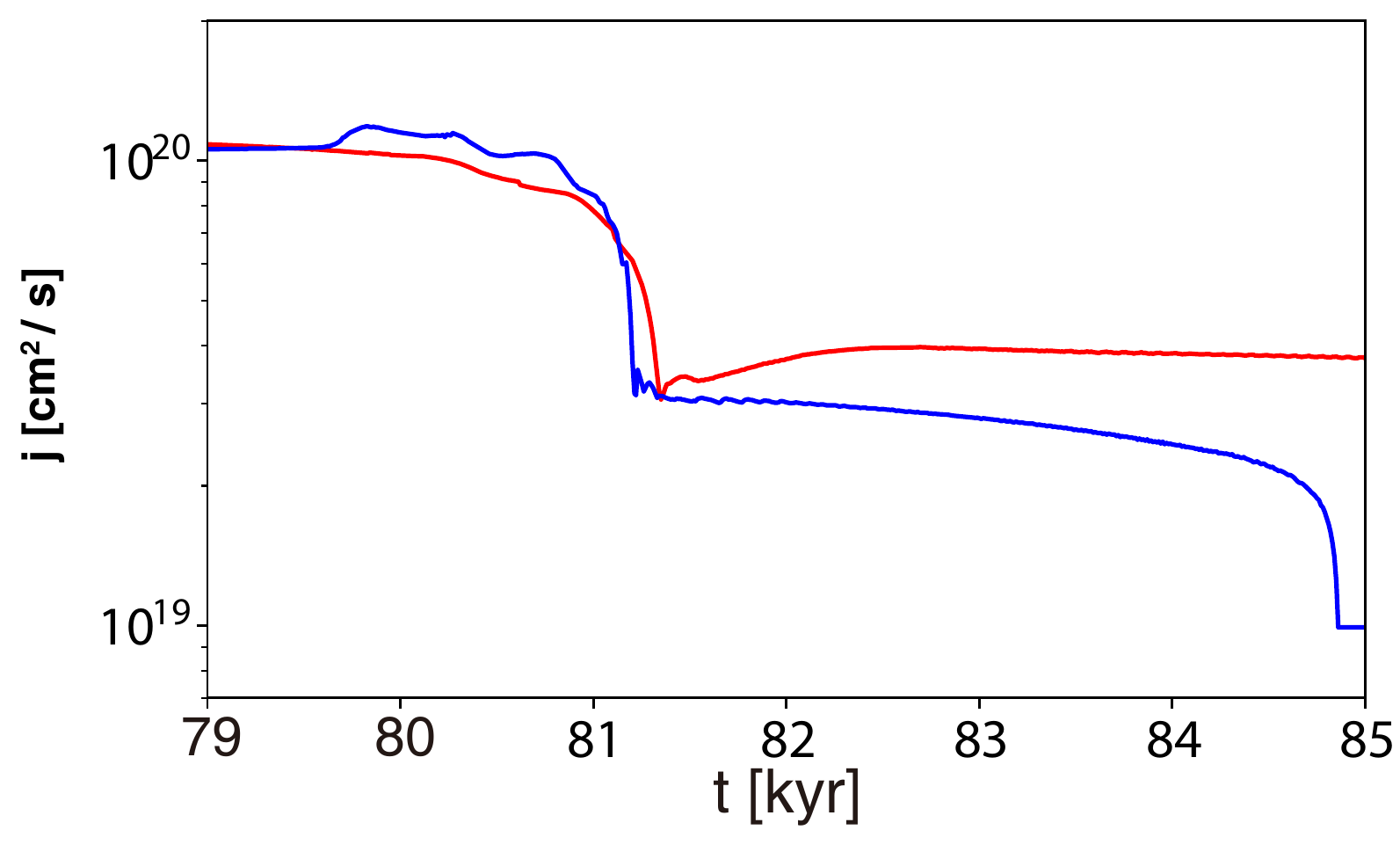}
    \caption{
Time evolution of specific angular momentum  for two selected dust particles. 
The red line corresponds to a dust particle orbiting around $10$\,au without showing significant inward radial movement, shown in Fig.~\ref{fig:l13_001mum_th90_r1500}. 
The blue line corresponds to the dust particle shown in Fig.~\ref{fig:l13_001mum_th45_r1900}, in which the particle falls onto the sink while orbiting in the disk. }
    \label{fig:tj001mum}
\end{figure}

To understand the  behavior of dust particles, the time evolution of the specific angular momentum for the two dust particles shown in Figs.~\ref{fig:l13_001mum_th90_r1500} and \ref{fig:l13_001mum_th45_r1900} is plotted in Fig.~\ref{fig:tj001mum}. 
In the figure, a sudden drop at $t\sim81$\,kyr indicates that the particles enter the disk from the infalling envelope. 
The dust particle plotted by the blue line gradually loses its specific angular momentum in the disk and finally falls onto the sink.  
The dust particle plotted by the red line increases its specific angular momentum during $t=81.2$--$82.5$\,kyr, and then  it has an almost constant specific angular momentum by the end of the simulation. 
Although the specific angular momenta of the two particles are almost the same before they enter the disk, their trajectories are different (see Figs.~\ref{fig:l13_001mum_th90_r1500} and \ref{fig:l13_001mum_th45_r1900}). 
The particle shown in Fig.~\ref{fig:l13_001mum_th90_r1500} enters the disk from the side on the equatorial plane, while the particle shown in Fig.~\ref{fig:l13_001mum_th45_r1900} enters the disk from above. 
The different trajectory histories in the envelope result in different outcomes.

\begin{figure}
    \centering
    \includegraphics[width=\linewidth]{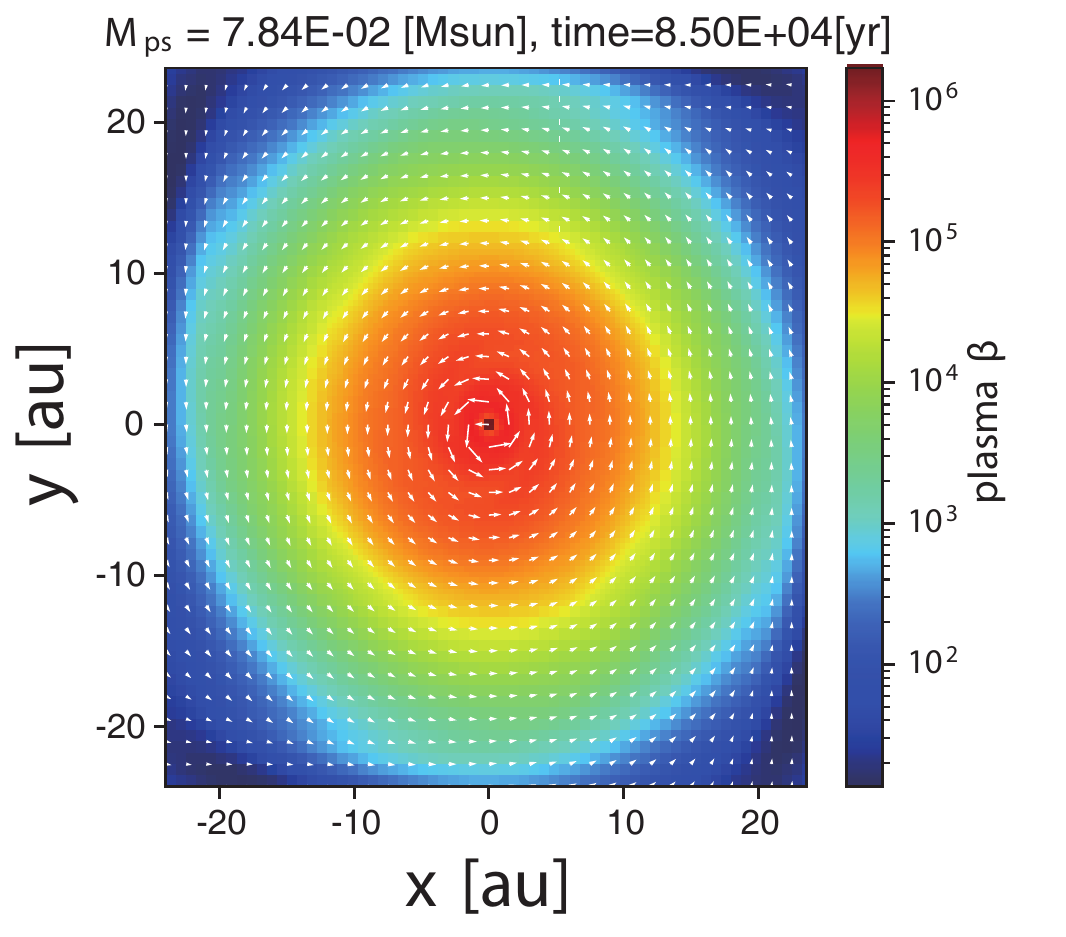}
    \caption{As Fig.~\ref{fig:l13q}, with the color indicating the plasma $\beta$.}
    \label{fig:beta}
\end{figure}

In addition to the gravitational torque, the angular momentum can be transported by magnetic effects. 
Plasma $\beta$ is a useful index to evaluate whether magnetic effects play a role in angular momentum transfer.
Fig.~\ref{fig:beta} plots plasma $\beta = P_{\rm gas} / P_{\rm mag}$ on the equatorial plane, where $P_{\rm gas}$ and $P_{\rm mag}$ are the gas  and magnetic pressure, respectively. 
The figure shows that the plasma beta around $r\sim10$\,au is as large as $\beta \sim 10^5$, indicating that magnetic effects can be negligible for transporting the angular momentum  in the rotationally supported disk.

\subsection{Difference from classical planet formation scenario}
\label{sec:scenario}
The MMSN is often used as a disk model in the classical planet formation scenario.
The gas disk considered in this study is different from the MMSN.
In this study, we simulated the formation and evolution of a disk from a prestellar cloud core (or a molecular cloud core).
Molecular cloud cores are  characterized by many observations \citep[e.g.,][]{2020ApJ...899...10T}.
We self-consistently include both the (disk) self-gravity and magnetic fields in our cloud core collapse  simulation. 
Our simulation shows a different dust behavior, which has not been reported in previous studies.
The crucial difference between  our study and previous studies is the radial inward movement of the dust. 
In previous studies, the dust gradually moves inward, which is called  the radial drift problem \citep[e.g.,][]{1977MNRAS.180...57W}. 
We show that a portion of the dust particles does not monotonically fall onto the central protostar, but that some particles can survive and orbit in the outer disk region. 
Many theoretical studies  have proposed possible solutions to avoid the radial drift problem.
In terms of a dust model, \citet{2012ApJ...752..106O} considered the internal structure of dust and showed that porous aggregates can solve the radial drift problem.
In addition, the radial drift problem can be overcome by preparing artificially adjusted gas disks. 
For example, \citet{2018ApJ...865..102T} suggested a gas disk model assuming an MHD disk wind producing a pressure bump within the disk.
In such a disk model,  dust grains concentrate around the pressure bump and  dust growth is significantly promoted before the dust falling onto the central protostar.

As described above, our simulation shows that a non-axisymmetric structure induced by gravitational instability develops in the disk  and  can prevent the radial drift of dust.
Our disk structure is obtained from a three-dimensional MHD simulation starting from the molecular cloud core and is suitable for investigating the initial conditions of planet formation. 
However, we ignore the dust growth and porosity.
The dust growth is discussed in the next section, while the porosity will be focused in future studies. 

\subsection{Dust Growth}
\label{sec:dustgrowth}
The growth of dust particles is ignored in our method, as described in \S\ref{sec:scenario} and Paper I.
Instead, we adopt different-size set of dust particles in the range $0.01\,\mu{\rm m} \le a_{\rm d} \le 1000\,\mu {\rm m}$. 
The simulation shows that the dust particles of any size remain in the outer disk region for $>3000$\,yr, as described in \S\ref{subsec:dustr}. 
Thus, the growth of dust particles can be expected in such a region (Fig.~\ref{fig:l13_001mum_th90_r1500}). 
This subsection discusses the possibility of the growth of dust particles orbiting around the outer disk region.

The evolution of dust peak (or typical) mass $m_{\rm d}$ in the Lagrange form can be described \citep{2014ApJ...789L..18O, 2016A&A...589A..15S, 2016ApJ...821...82O, 2021ApJ...920...27A} as 
\begin{equation}
\dfrac{dm_{\rm d}}{dt}=\dfrac{2\sqrt{\pi} a_{\rm d}^{2}\, \Delta v_{\rm d}\, \Sigma_{\rm d}}{h_{\rm d}},
\label{eq:dg1}
\end{equation}
where $\Sigma_{\rm d}$ and $h_{\rm d}$ are the surface density and scale height of the dust, and $\Delta v_{\rm d}$ is the relative velocity between dust particles. 
We consider the collisional growth of dust grains composed of single-sized particles with equation (\ref{eq:dg1}). 
As described in \S\ref{subsec:dustcouple}, the dust particles of any size adopted in this study are well coupled with the gas within a rotationally supported (or Keplerian) disk. 
Thus, the dust scale height $h_{\rm d}$ is assumed to be equal to the gas scale height $h_{\rm g}$.
In addition, the disk surface density can be described as $\Sigma_{\rm d}=f_{\rm dg} \Sigma_{\rm g}$, where $f_{\rm dg} (=0.01)$ is the dust-to-gas mass ratio. Note that the $f_{\rm dg}$ is considered not to be (significantly) changed during the simulation, as shown in Paper I. 
Then, with the gas mass density $\rho_{\rm g}$, we replace $(\Sigma_{\rm d}/h_{\rm d})$ into $ (f \Sigma_{\rm d}/h_{\rm d})=f\rho_{\rm g}$ in equation~(\ref{eq:dg1}).
Thus, equation~(\ref{eq:dg1}) can be written as 
\begin{equation}
\dfrac{dm_{\rm d}}{dt} = 2\sqrt{\pi} a_{\rm d}^2 \Delta v_{\rm d} f \rho_{\rm g}.
\label{eq:dg2}
\end{equation}
In addition, since we assume the spherical dust particles, the mass of each dust particle is described as 
\begin{equation}
m_{\rm d} = \dfrac{4\pi a_{\rm d}^3\, \rho_{\rm s}}{3},
\label{eq:dg3}
\end{equation}
where $\rho_{\rm s} (=1\,\gcm)$ is the material density of dust grains, in which the dust grains are assumed to be composed of ice (Paper I). 
With equations~(\ref{eq:dg1})--(\ref{eq:dg3}), we can estimate the growth timescale of dust grains $t_{\rm grow}$ with the dust mass $m_{\rm d}$ and  dust size $a_{\rm d}$ \citep{2021ApJ...920...27A,2021ApJ...920L..35T} as 
\begin{eqnarray}
t_{\rm grow} &\equiv& \left( \dfrac{1}{a_{\rm d}} \dfrac{da_{\rm d}}{dt}  \right)^{-1} 
= 3 \left( \dfrac{1}{m_{\rm d}} \dfrac{dm_{\rm d}}{dt} \right)^{-1} \nonumber  \\
&=& \dfrac{2\sqrt{\pi}a_{\rm d}\rho_{\rm s}}{\Delta v_{\rm d} f \rho_{\rm g}}. 
\label{eq:dg4}
\end{eqnarray}
We need the relative velocity between dust particles $\Delta v_{\rm d}$ to estimate the growth timescale in equation~(\ref{eq:dg4}).  
The relative velocity cannot be estimated in the simulation because the spatial scale between dust particles ($\lesssim 1$\,cm) and circumstellar disk ($>10^{14}$\,cm)  considerably differs. 
Thus, we use a simple formulation of the relative velocity  used in past studies \citep{2007A&A...466..413O, 2016ApJ...821...82O,2021ApJ...920...27A,2022arXiv221108947S} as
\begin{equation}
\Delta v_{\rm d} = \sqrt{3 \alpha\, {\rm St}}\, c_{\rm s},
\label{eq:dg5}
\end{equation}
where $\alpha$, St and $c_{\rm s}$ are the parameter characterizing turbulence or viscosity on a scale comparable to the size of dust particles, the Stokes number, and the sound speed of the outer disk region, respectively. 
Substituting equation~(\ref{eq:dg5}) into equation~(\ref{eq:dg4}), the growth timescale can be written as 
\begin{equation}
t_{\rm grow} = \dfrac{2\sqrt{\pi} a_{\rm d}\, \rho_{\rm s}}{f \rho_{\rm g}} (3\alpha\, {\rm St})^{-1/2} c_{\rm s}^{-1}.
\label{eq:dg6}
\end{equation}
In the following, we estimate the dust growth timescale (eq.~[\ref{eq:dg6}]) using our simulation results. 

As shown in Figures~\ref{fig:tr2d001mum}--\ref{fig:l13_001mum_th90_r1500}, the dust particles of any size remain (or continue to orbit) around $\sim 10$\,au far from the central protostar. 
In such a region, the gas density is as high as $\rho_{\rm g}\sim10^{-11}\,\gcm$ and the temperature is $T\simeq100$\,K (see Fig.~\ref{fig:l13_001mum_th90_r1500} and Paper I).  
Note that the $\rho_{\rm g}\sim10^{-11}\,\gcm$ roughly corresponds to the lowest density of the magnetically inactive region where the magnetic field dissipates \citep{2002ApJ...573..199N, 2007ApJ...670.1198M}. 
A rotationally supported disk forms in such a region in the early star formation stage \citep{2011MNRAS.413.2767M}. 
Next, using simulation results (Paper I and Figs.~\ref{fig:tSt_th90_r1000to2500_001mum_disk} and \ref{fig:tSt_th90_r2000to3500_1000mum_disk}), we relate the dust size $a_{\rm d}$ to the Stokes number St in the outer disk region ($\sim10$\,au) as 
\begin{equation}
{\rm St} \simeq 10^{-7} \left( \dfrac{a_{\rm d}}{10^{-6}\,{\rm m}} \right). 
\label{eq:dg65}
\end{equation}
Using these simulation results, we can rewrite  the growth timescale given by equation~(\ref{eq:dg6}) as 
\begin{equation}
t_{\rm grow} = 29.2 \left( \dfrac{a}{10^{-4}\,{\rm cm}} \right)^{1/2} \left( \dfrac{\rho_{\rm s}}{1\gcm} \right) 
\left( \dfrac{f}{0.01} \right)^{-1} \left( \dfrac{\rho_{\rm g}}{10^{-11}\,\gcm} \right)^{-1} 
 \left( \dfrac{\alpha}{0.01} \right)^{-1/2} \left( \dfrac{T}{100} \right)^{-1/2}  \ {\rm yr},   
\label{eq:dg66}
\end{equation}
where the sound speed $c_{\rm s}=7.0\times10^4 (T/100)^{1/2}$ is used. 
For simplicity, we describe the growth timescale as a function of only the dust size $a_{\rm d}$ as 
\begin{equation}
t_{\rm grow} = 29.2 \left( \dfrac{a}{10^{-6}\,{\rm m}} \right)^{1/2} \ {\rm yr}.
\label{eq:dg7}
\end{equation} 
Equation~(\ref{eq:dg7}) indicates that the dust grains with a size of $1\,\mu$m grow in $\sim30$\,yr in the outer disk region. 

As described in \S\ref{sec:results}, the dust particles entered from the horizontal direction (or the disk outer edge on the equatorial plane) orbit around $\sim10$\,au within the disk without falling onto the central protostar. 
The Keplerian timescale is given by 
\begin{equation}
t_{\rm Kep} = 100 \left(  \dfrac{M_*}{0.1M_\odot} \right)^{-1/2} \left( \dfrac{r}{10\,{\rm au}}  \right)^{3/2} \  {\rm yr}.
\end{equation}
The dust (and gas) particles in the outer disk region survive for $>3$\,kyr corresponding to $>30$ orbital periods at $r=10$\,au (see \S\ref{subsec:dustcouple}).  
Adopting the growth timescale being equal to $30\, t_{\rm Kep}=3$\,kyr, we can estimate the maximum dust size possible to be grown in the outer disk region with $t_{\rm grow}=30\, t_{\rm Kep}$. 
With equation~(\ref{eq:dg7}), we can estimate the maximum dust size as $a_{\rm d}=1.06$\,cm when the dust particle orbit around $\sim10$\,au for  $\sim3$\,kyr. 
In our calculation, the dust particle of any size ($0.01$--$1000\,\mu$m) are well coupled with the gas in the disk (\S\ref{subsec:dustcouple}). 
Thus, the growth of the dust particles should be justified until the dust size reaches $a_{\rm d}\simeq0.1-1$\,cm.

Recently, \citet{2021ApJ...920L..35T}, \citet{2022MNRAS.515.4780T} and \citet{2022MNRAS.514.2145B} investigated the dust growth in the early star formation stage with (magneto)hydrodynamic simulations.
The dust grows to reach  $a_{\rm d} \sim0.1-1$\,cm in \citet{2021ApJ...920L..35T} and \citet{2022MNRAS.515.4780T}, which is consistent with our estimate above. 
On the other hand, \citet{2022MNRAS.514.2145B} showed that dust grains only grow to about 0.01\,cm. 
However, since he investigated only the very early stage of star formation, his result does not contradict our estimate.

We stopped the calculation before the disk sufficiently grows due to the limitation of computational resources. 
Thus, the dust particles may grow larger than 1cm in size in a further evolutionary stage. 
Equation~(\ref{eq:dg65}) indicates that the $\rm{St} \le 0.1$ sustains as long as the dust size is smaller than $a_{\rm d}\lesssim1$\,m. 
Equation~(\ref{eq:dg7}) indicates that the dust can grow to reach $a_{\rm d}=1$\,m for about $3\times10^4$\,yr. 
When the dust grains have a size of $a_{\rm d}=1$\,m, the relative velocity is $\Delta v_{\rm d}=38$\,m\,s$^{-1}$ (eq.~\ref{eq:dg5}) and is still comparable to or smaller than the fragmentation velocity for the dust composed of ice grains $v_{\rm frag}=20-70$\,m\,s$^{-1}$ \citep[e.g.][]{2013A&A...559A..62W,2022MNRAS.515.2072K}.  
Thus, the dust may grow in the outer disk region until the dust size reaches $\sim1$\,m. 
Although we should miss many effects to suppress or promote the dust growth \citep{2022arXiv221013314B}, our result implies that the dust can sufficiently grow in the outer disk region. 
The rapid dust growth is attributed to the high-density and long dynamical timescale (with a small protostellar mass and large radius), which is realized in the early stage of star formation. 

\subsection{Observations of large-sized dust grains within protostellar outflow and envelope}
The dust growth in the disk would be verified  in the observations of protostellar outflows.  
\citet{2019ApJ...879...25K}, \citet{2019A&A...632A...5G} and \citet{2019MNRAS.488.4897V} have implied the existence of large-sized dust grains ($\gtrsim 10-100\,\mu$m) within the envelope. 
The dust growth timescale within the outflow is as long as $>10^8$\,yr, in which the maximum outflow density $\rho=10^{-17}\,\gcm$  (Fig.~\ref{fig:outflow}) and $a_{\rm d}=100\,\mu$m are introduced in equation~(\ref{eq:dg66}). 
In addition, the dust growth timescale in the envelope is $>10^6$\,yr with the density of $\rho=10^{-15}\,\gcm$.  
Thus, it is difficult for dust particles to grow to $\sim100\,\mu$m in the envelope and outflow. 

On the other hand,  we have already shown that the protostellar outflow contains $100\,\mu$\,m dust grains in our previous study  (see Figs.~8 and 12 of Paper I), indicating that the dust grains with a size of $10-100\,\mu$m are (partially) coupled with the gas within the outflow (see Fig.~13 of Paper I). 
Although we did not consider the dust growth in this and previous studies, we discuss the growth timescale of dust grains in \S\ref{sec:dustgrowth}. 
The dust grains can grow up to at least $\sim1$\,cm in size when the grains continue to orbit around the outer disk region (\S\ref{sec:dustgrowth}). 
Therefore, it is natural that dust grains with a size of $\gtrsim 10-100\,\mu$m are ejected from the circumstellar disk by the outflow.
\citet{2021ApJ...920L..35T} showed  that a portion of the dust particles are decoupled from the gas in the outflow and falls into the envelope and disk. 
As a result, the falling particles can contribute to the increase of the abundance of large-sized dust grains in the envelope. 
Thus, observing the large-sized grains in the envelope or outflow may be proof of dust growth in the circumstellar disk in the early stage of star formation. 
We cannot directly show the circulation and growth process of dust grains shown in \citet{2021ApJ...920L..35T}, because the dust growth is not considered in this study. 
We will investigate it in our future studies. 

\subsection{Caveats and future perspectives}
In addition to Ohmic dissipation, there are two other non-ideal MHD effects, ambipolar diffusion and  the Hall effect, which are not considered in our simulation.
Although these effects are also important during the star formation process,  they are not expected to significantly change  the dust trajectories in high-density gas disks, as both  ambipolar diffusion and  the Hall effect influence the gas dynamics in relatively low-density gas regions \citep[e.g.][]{2019MNRAS.484.2119K,2021MNRAS.504.5588K}.  
Our simulation settings are the same as in \citet{2017ApJ...835L..11T}, which reported that the simulation reproduces the observed disk structure fairly well. 
We show that dust particles located in the outer disk region do not fall onto the protostar because such particles receive angular momentum from the inner disk region, transported by the gravitational torque.  
The formation of a gravitationally unstable disk has also been confirmed with non-ideal MHD simulations that include three non-ideal MHD effects, Ohmic dissipation, ambipolar diffusion, and the Hall effect \citep[e.g.][]{2019MNRAS.489.1719W}.  
These simulations also show that angular momentum is transported outward by gravitational instability (or gravitational torque).
Thus,  suppression of the dust inward motion can be expected  as long as the disk is in a gravitationally unstable state or in the main accretion phase.

The magnetic diffusion coefficients of non-ideal MHD effects depend on the dust properties such as size and size distribution of dust grains \citep{2022MNRAS.515.2072K}.  
The non-ideal MHD effects or dust properties can influence the properties of protostellar outflows \citep[e.g.][]{2020ApJ...900..180M}. 
The outflow can transport angular momentum from the circumstellar disk. 
Thus, the change in the efficiency of angular momentum transport due to the outflow should change the disk properties, such as the size and density of the circumstellar disk in the early stage of star formation.
As a result, the properties of the region where dust particles continue to orbit should be changed.
However, the dust motion in the outer disk region should not be qualitatively changed as long as the gravitationally unstable disk sustains, as described above. 
On the other hand, the dust growth timescale discussed in \S\ref{sec:dustgrowth} may be changed according to the properties of both outflow and circumstellar disk.

Finally, we discuss the effect of radiative feedback from the protostar. 
In this study, we do not consider the issue of radiative transfer.
However, radiative heating from the central protostar may influence the disk dynamics.
The protostellar accretion luminosity can heat the disk even when the protostellar mass is as small as  $\sim0.1\,\msun$ \citep[e.g.][]{2013MNRAS.431.1719M, 2020ApJ...904..194H}.
Thus, radiative heating tends to  suppress the disk gravitational instability. 
In addition, viscous heating by differential rotation in the disk and shock heating during accretion from the infalling envelope  may affect the disk dynamics.
Heating and cooling should be carefully addressed in future studies to more adequately investigate the disk dynamics and behavior of dust grains.  

\section{Conclusion}
We investigated the behavior of dust grains in a star-forming cloud.
We developed a new method to calculate the behavior of dust grains as Lagrange particles and solve the fluid dynamics in the Eulerian framework to investigate the motion of dust grains in the early star and disk formation stage. We described the implementation and range of application for Lagrange dust particles in Paper I. 
We also showed the time evolution of dust abundance in the protostar, disk, outflow, and envelope regions in Paper I. 
The current paper focused on the dust motion and the coupling between dust and gas in the main accretion phase, during which the circumstellar disk gradually grows.

We prepared six different-size set of dust particles in the range $0.01$--$1000\,\mu$m and placed them in a  prestellar cloud core.
We calculated the evolution of the cloud core until the protostellar mass reaches 0.0784$\msun$, ignoring dust growth.  
In the gravitationally collapsing cloud, a circumstellar disk forms and drives a protostellar outflow. 
Although the disk gradually grows in the main accretion phase, the size of the disk is as small as $\sim20$\,au at the end of the simulation.

A large portion of the large dust particles ($a_{\rm d} \gtrsim 100\,\mu$m)  falls onto the disk, while the outflow can eject a small number. 
For the small dust particles ($a_{\rm d} \lesssim 10$--$100\,\mu$m), the dust particles initially placed with a small zenith angle $\theta_0 \lesssim 45^\circ$ are ejected by the outflow, where $\theta_0$ is the angle relative to the rotation axis (or the $z-$axis).  
Since we already showed the spatial variation in the abundance of different-sized dust grains in Paper I, we did not comment on this in this study. 

The motion of dust grains exhibits two trends after they enter the circumstellar disk, dependent on their history or trajectory until they reach the disk, regardless  of the dust size. 
The dust grains  slowly move inward and fall onto the sink (or protostar) when they enter the circumstellar disk from above, for which the incidence angle of the dust grains relative to the disk normal is small. 
In contrast, the dust grains continue to orbit near the outer edge of the disk, neither moving inward nor falling onto the protostar, when they enter the disk from the side or near the equatorial plane with a large incidence angle relative to the disk normal. 
In other words, dust grains reaching the inner disk region from the upper envelope rapidly fall onto the protostar, while those reaching the outer disk region survive without falling onto the protostar. 

We estimated the coupling between dust grains and  gas in the disk using the Stokes number to investigate the motion of the surviving dust grains. 
The Stokes numbers for all dust grains with sizes of $a_{\rm d}=0.01$--$1000\,\mu$m are well below unity, indicating that the dust grains are well coupled with the gas in the disk. 
Thus, the curious behavior of dust grains in the disk could be attributed to the gas motion within the disk.

In the disk, the magnetic field is weak and the plasma beta is very high because of effective magnetic dissipation.
Thus, the magnetic field cannot play a role in transporting the angular momentum in the disk.  
As a result, the disk mass gradually increases without an effective mechanism for angular momentum transfer.
This will cause gravitational instability because the mass supply from the infalling envelope to the disk persists in the main accretion phase. 
We found that a tiny spiral structure develops due to gravitational instability near the disk center.  
The spiral structure extends to $\sim10$\,au, while the disk has a radius of $\sim20$\,au. 
Thus, the gas and dust particles located at $r\lesssim10$\,au lose their angular momentum due to gravitational torque related to the spiral structure, and they can fall onto the central region. 
The gas and dust particles located at $r\gtrsim10$\,au receive angular momentum from the inner gas and dust particles, and survive without falling onto the protostar.

In our simulation, dust particles orbiting in the outer disk region survive for $>4000$\,yr without showing inward migration, 
but we cannot determine whether these particles survive for a further long period.  
However, if these particles survive for a very long time, the dust grains will grow and may form planetesimals or protoplanets. 
Thus, this study suggests that the outer disk regions where dust grains accumulate without inward radial drift could be a favored place of planet formation.
The number of dust particles calculated in this study is not large (102,984 particles, including gas test particles, for details, see Paper I), so it is difficult to quantitatively estimate the amount of dust grains orbiting in the outer disk region. 
The purpose of this study was to qualitatively investigate the dust motion in a growing disk during the main accretion phase and we do not extend the discussion to planet formation.
Future studies will investigate long-term integration to determine the final fate of dust grains orbiting in the outer disk region.

\section*{Acknowledgements}
We thank the referee for very useful comments and suggestions on this paper. 
We have benefited greatly from discussions with Yusuke Tsukamoto and Yoshihiro Kawasaki.
This work was supported by the Japan Society for the Promotion of Science KAKENHI (JP17H06360, JP17K05387, JP17KK0096, JP21H00046, JP21K03617: MNM), JSPS KAKENHI grants (numbers JP20J12062), and NAOJ ALMA Scientific Research Grant Code 2022-22B.
This research used the computational resources of the High-Performance Computing Infrastructure (HPCI) system provided by the CyberScience Center at Tohoku University, the Cybermedia Center at Osaka University, and the Earth Simulator at JAMSTEC through the HPCI System Research Project (project IDs hp200004, hp210004, hp220003). 
The simulations reported in this paper were also performed by 2021 and 2022 Koubo Kadai on the Earth Simulator (NEC SX-ACE and NEC SX-Aurora TSUBASA) at JAMSTEC.

\section*{Data Availability}
The data underlying this article are available in the article and in its online supplementary material.
 
\bibliographystyle{mnras}
\bibliography{koga} 

\bsp	
\label{lastpage}
\end{document}